\documentclass[aps,prd,showpacs,showkeys,preprintnumbers,epsf,nofootinbib]{revtex4}
\usepackage{graphicx,epsfig}
\newcommand{\bea}{\begin{eqnarray}}
\newcommand{\eea}{\end{eqnarray}}
\begin{document}
\preprint{arXiv: 1001.0493v4[hep-th]}
\title{A simple strategy for renormalization: QED at one-loop
level}
\author{Ji-Feng Yang$^{*,\dagger}$, Zhao-Ting Pan$^*$}
\address{$^*$Department of Physics, East China Normal University,
Shanghai 200241, China\\
$^\dagger$Kavli Institute for Theoretical Physics China, Chinese
Academy of Sciences, Beijing 100190, China}
%\date{April 28, 2010}
\begin{abstract}
We demonstrate our simple strategy for renormalization with QED at
one-loop level, basing on an elaboration of the effective field
theory philosophy. No artificial regularization or deformation of
the original theory is introduced here and hence no manipulation of
infinities, ambiguities arise instead of infinities. Ward identities
first come to reduce the number of ambiguities, the residual ones
could in principle be removed by imposing physical boundary
conditions. Renormalization group equations arise as "decoupling
theorems" in the underlying theory perspective. In addition, a
technical theorem concerning routing of external momenta is also
presented and illustrated with the self-energy and vertex function
as examples.
\end{abstract}
\maketitle% \tableofcontents
\section{Introduction}
All the presently known quantum field theories for practical use are
beset with ultraviolet divergences to certain extent in perturbative
frameworks. These divergences imply that such theories are not well
defined at short distances before renormalization is carried out, at
least within perturbative frameworks. The celebrated BPHZ program is
the standard algorithm for renormalization\cite{BPHZ1,BPHZ2,BPHZ3}
that is widely employed by particle and field theorists, which can
lead to a mathematically well-defined perturbative formulation of a
field theory in the end. But the procedures involved are not easy to
be endowed with a natural logic that would be readily to accept. In
this connection, Epstein and Glaser (EG) had already shown that
finite perturbation theory of quantum field theories can be obtained
through constructive methods basing on causality
conditions\cite{EG1,EG2} where Feynman rules are only valid at tree
level, which provides more mathematical rationales for conventional
procedures like BPHZ\cite{EGBP1,EGBP2}. Recently, other mathematical
structures hidden in the perturbative formulation (especially in
BPHZ) of quantum field theories has become a rapidly growing area of
research\cite{Hopf1,Hopf2,Hopf3}, which implies that there are more
regularities worth exploring under the veil of renormalization.

From more physical point of view, according to the effective theory
philosophy, the appearance of ultraviolet divergences can be
understood from the fact that the presently known formulation of a
quantum field theory is necessarily an effective one with the
sophisticated structures or parameters dominant at short distances
or larger energy scales being ignored. Unfortunately, the
sophisticated structures underlying the present formulation of
quantum field theories are actually unavailable or not discovered
yet. Nor could us simply guess about the exact answers. The
divergences or the ill-defined objects are the prices we paid for
such forced ignorance or simplification. This is the start of the
effective field theory philosophy that is now widely accepted and
applied by physicists. Fortunately, this effective field theory view
also implies that the ignored short-distance details could at best
affect the physical behaviors of effective field theories through
local interactions or operators that could be parametrized as
corrections to the actions of quantum field theories, which, after
being absorbed into the normalization factors for effective field
theory's couplings and operators, leads to the {\em renormalization}
of quantum (effective) field theories. Otherwise they would be
"captured" as explicit and active degrees of freedom.

In practical computations, one usually has to first introduce
various regularization schemes and intermediate divergences to be
removed later, which are supposed to imitate the roles played by the
true underlying structures or parameters. Actually, with such
operations, the perturbative expansion only takes rather formal
meaning due to the presence of infinite counter-terms. Moreover, for
the regularization schemes or deformations to be legitimate and
useful, they must not harm the bulk structures of the original
theory as well as being self-consistent. In particular, important
symmetries and global properties like the topologies of spacetime
and Hilbert space, unitarity and causality, etc., which serve as
foundations for a quantum field theory, should be preserved as much
as possible, at least no drastic or uncontrollable changes should be
introduced. However, the status of the principles just stated are
extremely hard to precisely or profoundly assess within any known
regularization. As a matter of fact, the comprehensive and global
structures of a well-defined quantum field theory have never been
fully delineated, and we heavily rely on rather formal evaluations
to deal with this very difficult issue. Sometimes, to pursue
efficiency, even the logical foundations of the theory would be
severely mutilated\cite{visser}. Thus, we only accumulate our faiths
in one regularization method through intensive tests against
experiment data. Up till now, there exists no regularization that
could be satisfactorily in service in all contexts, each has certain
shortcomings that will fail somewhere, or alternatively, no one
could prove to work everywhere\footnote{The reason is obvious: Were
it found, it must at least be equivalent to the true underlying
theory that is well-defined in every aspect, including the correct
formulation of quantum gravity.}. For example, the celebrated
dimensional regularization works excellently for standard model,
especially for the strong interaction, but not for supersymmetric
theories at
all\cite{thooft1,thooft2,thooft3,thooft4,thooft5,thooft6,thooft7,thooft8}.
Of course, other schemes also have their own shortcomings, for a
comprehensive discussions of the advantages and disadvantages of
various existing regularization methods, please refer to
Ref.\cite{WuYL}. This is also an indirect reflection of our
ignorance of the grand structures or contents of quantum field
theories in depth. Therefore, most regularization schemes are
actually used in a superficial or formal sense, only appearing in
the intermediate stages of calculations. After the artificial and
transient excursion to parametrize our ignorance of the underlying
short-distance physics, we "return" to our "original" quantum field
theories with the deformations removed somehow. This practice is
quite established within perturbative regimes for many physically
interested quantities or operators. Nevertheless, the ultimate
theoretical values for such doings still remain to be seen, at least
in nonperturbative regimes.

On account of the above reasonings, it is definitely desirable or
worthwhile to explore of any approach or procedure that would
introduce no or as less as possible deformation of the bulk aspects
of a quantum field theory and the associated infinities or
divergences. For physicists, it is also more convenient to work with
Feynman rules and Feynman diagrams that are physically intuitive and
well defined at tree level in such a manner that the radiative
corrections or loop diagrams could be computed without introducing
too much mathematical auxiliaries. Actually, as will be demonstrated
and argued below, there may exist an approach or strategy that seems
to be free of the problems mentioned above or satisfy the
requirements just enumerated. It is actually based on an elaboration
of the effective field theory
philosophy\cite{talk10,talk11,talk2,981,982}, introducing no
artificial deformations, and more pleasantly, no ultraviolet
infinities and the associated subtraction of infinities. Only local
ambiguities for each loop integral may occur as a parametrization of
our ignorance, which, as in the last step of renormalization, are to
be fixed through appropriate boundary conditions. We feel all these
virtues should make our strategy a simple and natural approach to
start with. It has already been partially employed in a number of
field theoretical issues\cite{Acta951,Acta952,PRD65,PRC71,SciChn}.

From this report on, we wish to launch serial works to demonstrate,
explain and further develop our strategy, with our ultimate goals
targeting at a consistent and efficient program. In the course of
these works, we will illustrate through concrete examples that many
technical subtleties such as external momenta routing (or shift of
integration variables) and overlapping divergent diagrams do not
cause any trouble. From our demonstrations, one could find that our
strategy could be well applied beyond the standard perturbative
framework in terms of Feynman diagrams, see the applications in
nonperturbative problems\cite{PRC71,SciChn}, unlike those
principally devised for and limited to relativistic perturbative
formulation. Another virtue of our approach is that we may examine
important issues such as symmetry status of the radiative
corrections (perturbative or nonperturbative) without being stuck to
some specific prescription that might lead us to wrong conclusions
or judgement\cite{Acta951,Acta952,PRD65}. For perturbative
amplitudes, our strategy could reproduce the EG approach's results.
In fact, our physically motivated strategy share the following
consensus with the mathematically motivated EG approach: the
perturbative quantum field theories are not unambiguously defined in
the short distances, some elementary Feynman or loop diagrams
contain local ambiguous terms that could only be fixed through
appropriate boundary conditions. In this report, we employ the
simple and well-known objects in quantum electrodynamics (QED) for a
clear and pedagogical illustration of the "naturalness" and
simplicity of the principles and perspective that are adopted in our
approach. In a sense, we wish to motivate an elaboration of the
effective field theories philosophy with respect to renormalization
of all quantum processes.

The organization of this report is as follows: We describe our
strategy in Sec. II along with some technical issues. In Sec. III,
we compute the elementary one-loop amplitudes or Feynman graphs
(that are ill defined in QED) using natural differential equations
following from the existence of a complete underlying theory. The
Ward identities between the vertex and self-energy diagrams will be
demonstrated as a valid constraint and the four-photon (light by
light scattering) diagrams will be shown to be finite as an
additional consequence of gauge invariance. In Sec. IV, we consider
how to impose appropriate boundary conditions to fix the
ambiguities, how this step is linked to the conventional procedures,
and related issues. We will also answer the origin of
renormalization group equations in Sec. IV. The whole presentation
is summarized in Sec. V, where some conceptual issues will also be
addressed.
\section{Descriptions of our strategy}
Our strategy is in fact to determine the ill-defined (or undefined)
loop amplitudes through solving natural differential equations
together with imposition of physical boundary conditions. In terms
of Feynman diagrams, this is to "calculate" divergent diagrams using
convergent ones. To proceed, we denote a superficially divergent or
ill-defined diagram and its underlying theory version with the same
symbol $\Gamma$ which should be a function of external momenta
$[p]$, couplings and masses $[g]$ and the underlying parameters
$\{\sigma\}$ (which will be usually hidden to avoid heavy formulae)
that render the diagram finite or mathematically meaningful. As the
underlying parameters $\{\sigma\}$ must be very small in "size", an
ill-defined diagram should be determined through the "decoupling"
limit of the version with the underlying parameters. Not knowing the
details of the underlying theory, a natural and yet legitimate
operation about the underlying theory version of an ill-defined
diagram is to differentiate it with respect to external momenta
$[p]$ for enough times so that the resulting loop integration could
be computed in terms of the conventional Feynman rules of quantum
field theories\footnote{In underlying theory perspective, the loop
integration and "decoupling" limit (or low-energy projection)
commute with each other on such differentiated diagrams, or that the
"decoupling" limit can be performed before loop integration is
done\cite{talk10,talk11,talk2}.}:\bea\partial^{\omega_{\Gamma}+1}_p
\Gamma([p],[g])=\tilde{\Gamma}^{(\omega_\Gamma+1)}([p],[g]),\eea
with $\omega_{\Gamma}$ being the superficial divergence degree of
diagram $\Gamma$ and $\tilde{\Gamma}^{(\omega_\Gamma+1)}$ denoting
the well-defined diagrams generated by the operation
$\partial^{\omega_{\Gamma}+1}_p$. Then we solve the above
differential equation to arrive at a finite expression of $\Gamma$
that contains a polynomial in terms of external momenta $[p]$ with
ambiguous coefficients. So, the general solution for the ill-defined
diagram $\Gamma$ reads:\bea \Gamma([p],[g])=\tilde{\Gamma}([p],[g])
+\Gamma^{\text{Poly}}_{[\omega_{\Gamma}]}([p],[g],[c]).\eea Here,
$[c]$ denotes the integration constants collectively. As last step
of our strategy, the ambiguities are removed through imposing
reasonable symmetries AND appropriate boundary
conditions\cite{Ster}. It is obvious that in our strategy, {\em no
artificial deformation is introduced at all}.

As our method rely on operations with respect to external momenta,
there may be concerns about the issue of external momenta routing.
Below we will show that:

\newtheorem{rou}{Theorem of Routing}[section]\begin{rou}
The routing of external momenta in a loop diagram does not matter at
all in our strategy, i.e., different routings will at most yield a
reparametrization of the ambiguous polynomials.\end{rou}

The routing of external momenta has been first addressed in
Refs.\cite{981} and \cite{982} more than a decade ago. Here, we give
a more concrete formulation of the remarks given there. There may be
two kind of routings: (a) one keeps the distribution of external
momenta at vertices intact; (b) the other amounts to relabeling the
external momenta at vertices: $[p]\rightarrow[p^\prime]$, or a
redefinition of the external momenta at vertices. First, we prove it
for one-loop amplitudes.

\textbf{Proof. Case (a)}: As the underlying theory version is
well-defined, one could perform a integration variable
transformation so that a different routing is exactly resulted,
thus,\bea\label{routinga}
\Gamma([p],[g];\{\sigma\})_{\text{rt1}}=\Gamma([p],[g];\{\sigma\}
)_{\text{rt2}},\eea with the underlying parameters $\{\sigma\}$
being now explicitly included for illustration. That is, this
variable transformation should not alter the evaluation. Now apply
the differentiation with respect to the external momenta
$\partial^{\omega_\Gamma}_{[p]}$ on Eq.(\ref{routinga}), we end up
with the following equality: \bea
\partial^{\omega_\Gamma}_{[p]}\Gamma([p],[g];\{\sigma\})_{\text{rt1}}
=\partial^{\omega_\Gamma}_{[p]}\Gamma([p],[g];\{\sigma\})_
{\text{rt2}}.\eea Taking the "low-energy" limit so that $\{\sigma\}$
become "decoupled", we have:\bea\text{Limit}_{\{\sigma\}}&&\left\{
\partial^{\omega_\Gamma}_{[p]}\left[\Gamma([p],[g];\{\sigma\}
)_{\text{rt1}}-\Gamma([p],[g];\{\sigma\})_{\text{rt2}}\right]
\right\}\nonumber\\=&&\partial^{\omega_\Gamma}_{[p]}
\left[\Gamma([p],[g])_{\text{rt1}}-\Gamma ([p],[g])_
{\text{rt2}}\right]=0.\eea That means different routings differ at
most by an polynomial $\Gamma^{\text{Poly}}$ that would be
annihilated by $\partial^{\omega_\Gamma}_{[p]}$:\bea
\Gamma^{\text{Poly}}([p],[g])\equiv\Gamma([p],[g])_{\text{rt1}}
-\Gamma([p],[g])_{\text{rt2}},\eea which could well be absorbed into
the ambiguous polynomials in $\Gamma$ (to be fixed later through
physical boundaries) as a reparametrization of the expression.

\textbf{Case (b)}: In this case the operation about external momenta
should also not alter the amplitude provided the integration is done
in underlying theory:\bea\label{routingb}\Gamma([p],[g];\{\sigma\})
=\Gamma ([p^\prime],[g];\{\sigma\}).\eea Now apply the
differentiation with respect to the external momenta $[p]_0$ that
are in a subset shared by $[p]$ and $[p^\prime]$ to
Eq.(\ref{routingb}), we have,\bea
\partial^{\omega_\Gamma}_{[p]_0}[\Gamma([p],[g];\{\sigma\})
-\Gamma([p^\prime],[g];\{\sigma\})]=0.\eea Then we can take the
"decoupling" limit and perform the loop integrations in the two
routing, just as in case (a). Again, the net difference between the
two routings is at most a polynomial with respect to $[p]_0$ that
could again be absorbed into the ambiguous part, after we return to
the original label of external momenta, the same expression for
$\Gamma$ should be restored.

The multi-loop diagrams could be treated in similar
fashion\cite{talk10,talk11,talk2,981,982}: First, differentiate the
diagram with respect to the momenta that are external to the overall
diagram to annihilate the overall divergences; Second, consider each
individual sub-diagram that is divergent and differentiate it with
respect to the momenta that are {\em external to this sub-diagram}
to annihilate the "overall" divergences of the sub-diagram; Third,
continue this operation till the smallest sub-diagrams are thus
differentiated; Fourth, carry out all the resulting loop
integrations and integrate back indefinitely with respect to the
corresponding momenta that are external to the corresponding loops;
The final outcome will be a finite expressions in terms of momenta
and masses external to the whole diagram, containing ambiguities
associated each individual divergent or ill-defined loops. This
operation naturally dissolves the overlapping divergences that are
notoriously difficult to deal with in conventional treatments,
thanks to the work by Caswell and Kennedy\cite{caswell}. Since the
treatment should be done loop by loop, thus, the theorem of routing
applies to each loop integration as well and finally to the whole
diagram. {\em Q.E.D.}

This theorem would allow us to choose any routing we like or that is
convenient. Obviously, simple routing of external momenta would
yield simplicity in calculations.
\section{QED at one-loop level} We will work with the following
standard covariant gauge QED in 3+1-dimensional spacetime:
\bea\label{QEDlag}{\mathcal{L}}_{\text{QED}}={\bar\psi}
(i\slash\!\!\!\!D-m)\psi-\frac{1}{4}F_{\mu\nu}F^{\mu\nu}
-\frac{1}{2\xi}(\partial_\mu A^\mu)^2,\eea with $\xi$ being the
dimensionless gauge parameter. Our metric convention reads \bea
g_{\mu\nu}=g^{\mu\nu}=\left(+,-,-,-\right).\eea

At one-loop level, only the following five elementary vertex
functions are superficially UV divergent and hence have to be
renormalized: the self-energy $\Sigma$, the electron-photon vertex
$\Lambda_\mu$, the vacuum polarization $\Pi_{\mu\nu}$, the three-
and four-point photon vertices. The Feynman diagrams for the first
three vertices are given in Fig.1. The three-point photon vertex is
zero due to Furry's theorem, while, as will be shown below, the four
point photon vertex is actually definite in our approach due to
gauge invariance.

We first consider the vertex functions listed in Fig.1 whose Feynman
integrals read,\bea&&-i\Sigma^{(1)}(p)\equiv(-ie)^2\int\frac{d^4l}
{(2\pi)^4}\frac{i\left[(1-\xi)\frac{l^\kappa l^\tau}{l^2}
-g^{\kappa\tau}\right]}{l^2+i\epsilon}\gamma_\kappa \frac{i}
{\slash\!\!l+\slash\!\!\!p-m+i\epsilon}\gamma_\tau;\\
&&\Lambda^{(1)}_\mu(p,\tilde{p})\equiv(-ie)^2\int\frac{d^4l}
{(2\pi)^4}\frac{i\left[(1-\xi)\frac{l^\kappa l^\tau}{l^2}
-g^{\kappa\tau}\right]}{l^2+i\epsilon}\gamma_\kappa\frac{i}
{\slash\!\!l+\slash\!\!\!p-m+i\epsilon}\nonumber\\&&\quad\quad\quad
\quad\quad\quad\times\gamma_\mu\frac{i}{\slash\!\!l
+\slash\!\!\!\tilde{p}-m+i\epsilon}\gamma_\tau;\\
&&i\Pi^{(1)}_{\mu\nu}(q)\equiv(-ie)^2\int\frac{d^4l}{(2\pi)^4}(-1)
\texttt{Tr}\left[\gamma_\mu\frac{i}{\slash\!\!l-m+i\epsilon}
\gamma_\nu\frac{i}{\slash\!\!l+\slash\!\!\!q-m+i\epsilon}\right],
\eea where $p,\tilde{p} $ and $q$ denote the external momenta for
electron and photon respectively. The superficial divergence degrees
for the three diagrams are well-known: $\omega(\Sigma^{(1)})=1$,
$\omega(\Lambda^{(1)}_\mu)=0$, $\omega(\Pi^{(1)}_{\mu\nu})=2$. Note
the routing of the external momenta are so chosen that they flow
through the fermionic internal lines, which will yield convenience
for calculations and a clear diagrammatic interpretation of the
operation associated with the differentiation with respect to
external momenta, see below. We will consider other routings for the
self-energy and the vertex diagrams in the Appendix A.
\subsection{Differential equations for divergent integrals} Below,
we treat these superficially divergent integrals with the strategy
described above, according to which the three differential equations
are,\bea\label{diffsf}&&\partial_{p_{\alpha}}\partial_{p_{\beta}}
\Sigma^{(1)}(p)=\Sigma^{(1);\alpha\beta}(p),\\
\label{diffvt}&&\partial_{p_{\alpha}}\Lambda^{(1)}_\mu(p,
\tilde{p})=\Lambda^{(1);\alpha}_\mu(p,\tilde{p}),\\
\label{diffvp}&&\partial_{q_{\alpha}}\partial_{q_{\beta}}
\partial_{q_{\gamma}}\Pi^{(1)}_{\mu\nu}(q)=\Pi^{(1);\alpha\beta
\gamma}_{\mu\nu}(q),\eea where the right hand sides (RHS) of these
differential equations are convergent Feynman diagrams, their
corresponding loop integrals are listed in Appendix A. Actually, in
terms of Feynman diagram the RHS of
Eqs.(\ref{diffsf},\ref{diffvt},\ref{diffvp}) are just obtained
through insertion of photon probes with zero momentum, provided that
the external momentum flows through internal fermion lines, cf.
Fig.2 for an example.

The loop integration on the RHS of
Eqs.(\ref{diffsf},\ref{diffvt},\ref{diffvp}) can now be
straightforwardly carried out:\bea-i\Sigma^{(1);\alpha\beta}(p)=
&&\frac{8ie^2}{(4\pi)^2}\int^1_0\!\!\!z^3dz\left\{\frac{(z\slash
\!\!\!p-2m)p^\alpha p^\beta}{z(zp^2-m^2)^2}-\frac{g^{\alpha\beta}
(\slash\!\!\!p-\frac{2m}{z})+\gamma^{\{\alpha}p^{\beta\}}}
{2z(zp^2-m^2)}\nonumber\right.\\&&\left.+(1-\xi)\left[\frac{
(p^2-m^2)\slash\!\!\!pp^\alpha p^\beta}{(zp^2-m^2)^3}-\frac{
(p^2-m^2)(g^{\alpha\beta}\slash\!\!\!p+\gamma^{\{\alpha}p^{\beta\}}
)}{4z(zp^2-m^2)^2}\right.\right.\nonumber\\&&\left.\left.-\frac{
(6\slash\!\!\!p-2m)p^\alpha p^\beta}{4z(zp^2-m^2)^2}+\frac{
g^{\alpha\beta}(\slash\!\!\!p-\frac{m}{2})+\gamma^{\{\alpha}
p^{\beta\}}}{2z^2(zp^2-m^2)}\right]\right\},\\
\Lambda^{(1);\alpha}_\mu(p,\tilde{p})=&&-\frac{e^2}{(4\pi)^2}
\int_0^1\!\!\!dy\int_0^{1-y}\!\!\!dz\left\{-\frac{4z\gamma_\mu
(p-P)^\alpha}{\Delta}+\frac{2z(p-P)^\alpha\mathcal{G}_\mu}{\Delta^2}
\right.\nonumber\\&&\left.+\gamma^\sigma\frac{(1-z)\gamma^\alpha
\gamma_\mu (\slash\!\!\!\tilde{p}+m-\slash\!\!\!P)-z(\slash\!\!\!p
+m -\slash\!\!\!P)\gamma_\mu\gamma^\alpha}{\Delta}\gamma_\sigma
\right.\nonumber\\&&\left.+(1-\xi)(1-y-z)\left[\frac{4z(p-P)^\alpha
\mathcal{F}_{1;\mu} }{\Delta^3}\right.\right.\nonumber\\&&\left.
\left.+\frac{[z\gamma^\alpha(\slash\!\!\!p+m-\slash\!\!\!P) +(1-z)
\slash\!\!\!P\gamma^\alpha]\gamma_\mu(\slash\!\!\!\tilde{p}
+m-\slash\!\!\!P)\slash\!\!\!P}{\Delta^2}\right.\right.\nonumber\\&&
\left.\left.+\frac{z\slash\!\!\!P(\slash\!\!\!p+m -\slash\!\!\!P)
\gamma_\mu[(\slash\!\!\!\tilde{p} +m-\slash\!\!\!P)\gamma^\alpha
-\gamma^\alpha \slash\!\!\!P]}{\Delta^2}\right.\right.\nonumber\\&&
\left.\left.+\frac{2z(p-P)^\alpha\mathcal{F}_{2;\mu}}{\Delta^2}
+\frac{3z\gamma_\mu \slash\!\!\!\tilde{p}\gamma^\alpha +3z\gamma^
\alpha\slash\!\!\!p\gamma_\mu+3\slash\!\!\!P\gamma^\alpha\gamma_\mu}
{\Delta}\right.\right.\nonumber\\&& \left.\left.+\frac{\slash
\!\!\!\tilde{p}\gamma_\mu\gamma^\alpha +12z(p-2P)^\alpha\gamma_\mu
+(6z-2)m g^\alpha_\mu}{\Delta}\right]\right\},\\
i\Pi^{(1);\alpha\beta\gamma}_{\mu\nu}(q)=&&\frac{2^4ie^2}{(4\pi)^2}
\int^1_0\!\!\!z^4(1-z)^4dz\left\{\frac{g_{\mu}^{\{\alpha}q^\beta
g^{\gamma\}}_{\nu}+q_{\{\mu}g_{\nu\}}^{\{\alpha}g^{\beta\gamma\}}
-2g_{\mu\nu}g^{\{\alpha\beta}q^{\gamma\}}}{z^2(1-z)^2(m^2-z(1-z)q^2)}
\right.\nonumber\\&&\left.+\frac{q_{\{\mu}g_{\nu\}}^{\{\alpha}
q^{\beta}q^{\gamma\}}-12g_{\mu\nu}q^{\alpha}q^\beta q^{\gamma}
+(q_\mu q_\nu-p^2g_{\mu\nu})g^{\{\alpha\beta}q^{\gamma\}}}{z(1-z)
(m^2-z(1-z)q^2)^2}\right.\nonumber\\&&\left.+\frac{8(q_\mu q_\nu
-q^2g_{\mu\nu})q^{\alpha}q^{\beta}q^{\gamma}}{(m^2-z(1-z)q^2)^3}
\right\},\eea\bea P\equiv&&zp+y\tilde{p},\ \Delta\equiv(y+z)m^2+P^2
-zp^2-y\tilde{p}^2,\nonumber\\\mathcal{G}_\mu\equiv&&\gamma^\sigma
(\slash\!\!\!p+m-\slash\!\!\!P)\gamma_\mu(\slash\!\!\!\tilde{p}
+m-\slash\!\!\!P)\gamma_\sigma ,\nonumber\\
\mathcal{F}_{1;\mu}\equiv&&\slash \!\!\!P(\slash\!\!\!p+m-\slash
\!\!\!P)\gamma_\mu(\slash \!\!\!\tilde{p}+m-\slash\!\!\!P)\slash
\!\!\!P,\nonumber\\ \mathcal{F}_{2;\mu}\equiv&&\slash\!\!\!\tilde{p}
\gamma_\mu\slash\!\!\!p+3\gamma_\mu\slash\!\!\!\tilde{p}\slash
\!\!\!P +3\slash\!\!\!P\slash\!\!\!p\gamma_\mu+(m^2-6P^2)\gamma_\mu
+2m(3P_\mu-p_\mu-\tilde{p}_\mu).\eea Here symmetrization about Greek
indices grouped in the bracket $\{\cdots\}$) is frequently used for
brevity. The three definite functions then determine the original
loop amplitudes up to certain degree of ambiguity. Next, we proceed
to solving the differential equations using the above definite
functions.
\subsection{Solutions}
Evidently, the direct integrating back with respect to external
momenta would be cumbersome. Instead of doing so, we may first
extract the differentiation with respect to external momenta in the
following manner,\bea-i\Sigma^{(1);\alpha\beta}(p)=&&\partial_{
p_{\alpha}}\partial_{p_{\beta}}\left\{\frac{ie^2}{(4\pi)^2}\int_0^1
\!\!\!dz\left[[(1-2z-\xi)\slash\!\!\!p+(3+\xi)m]\ln(m^2-zp^2)
\right.\right.\nonumber\\&&\left.\left.-(1-\xi)\frac{z(p^2-m^2)
\slash\!\!\!p}{m^2-zp^2}\right]\right\},\\
\Lambda^{(1);\alpha}_\mu(p,\tilde{p})=&&\partial_{p_\alpha}\left\{
-\frac{e^2}{(4\pi)^2}\int_0^1\!\!\!dy\int_0^{1-y}\!\!\!dz\left[2
\gamma_\mu\ln\Delta+\frac{\mathcal{G}_\mu}{\Delta}+(1-\xi)(1-y-z)
\right.\right.\nonumber\\&&\left.\left.\times\left(-6\gamma_\mu
\ln\Delta+\frac{\mathcal{F}_{1;\mu}}{\Delta^2}+\frac{
\mathcal{F}_{2;\mu}}{\Delta}\right)\right]\right\},\\
i\Pi^{(1);\alpha\beta\gamma}_{\mu\nu}(q)=&&\partial_{q_{\alpha}}
\partial_{q_{\beta}}\partial_{q_{\gamma}}\left\{-\frac{8ie^2}
{(4\pi)^2}(q_\mu q_\nu-g_{\mu\nu}q^2)\int^1_0\!\!\!dzz(1-z)
\right.\nonumber\\&&\left.\times \ln(m^2-z(1-z)q^2)\right\},\eea
then the solutions could be readily found as:\bea\label{solutionsf}
-i\Sigma^{(1)}(p)&&=\frac{ie^2}{(4\pi)^2}\left\{\int_0^1\!\!\!dz
\left[[(1-2z-\xi)\slash\!\!\!p+(3+\xi)m]\ln(m^2-zp^2)\right.\right.
\nonumber\\&&\quad\left.\left.-\frac{(1-\xi)z(p^2-m^2)
\slash\!\!\!p}{m^2-zp^2}\right]+C_\psi\slash\!\!\!p+C_m\right\},\\
\label{solutionvt}\Lambda^{(1)}_\mu(p,\tilde{p})&&=-\frac{e^2}
{(4\pi)^2}\left\{\int_0^1\!\!\!dy\int_0^{1-y\!\!\!}dz\left[2
\gamma_\mu\ln\Delta+\frac{\mathcal{G}_\mu}{\Delta}+(1-\xi)(1-y-z)
\right.\right.\nonumber\\&&\left.\left.\times\left(-6\gamma_\mu
\ln\Delta+\frac{\mathcal{F}_{1;\mu}}{\Delta^2}+\frac{\mathcal{F}_{2;\mu}}
{\Delta}\right)\right]+C_\perp\gamma_\mu\right\},\\
\label{solutionvp}i\Pi^{(1)}_{\mu\nu}(q)&&=-\frac{8ie^2}{(4\pi)^2}
\left[(q_\mu q_\nu-g_{\mu\nu}q^2)\int^1_0\!\!\!dzz(1-z)
\ln(m^2-z(1-z)q^2)\right.\nonumber\\&&\quad\left.+C_Aq_\mu q_\nu
+\tilde{C}_{A}g_{\mu\nu}q^2+C_{\{\mu}q_{\nu\}}+C_{\mu\nu}
+g_{\mu\nu}C_{\Pi;0}\right],\eea with $[C_{\cdots}]$ ($C_{\mu\nu}$
is traceless, $C^\mu_\mu=0$) being the corresponding integration
constants that are ambiguous at present stage. They are uniquely
defined only in the underlying formulation that is well defined at
short distances, but in the effective theory, i.e., QED, they are
free constants to be determined through appropriate boundary
conditions\cite{Ster}.

Actually, for photon vacuum polarization, Lorentz invariance could
remove the constants $C_\mu,C_{\mu\nu}$. The presence of $C_{\Pi;0}$
would violate gauge invariance, i.e, making the photon massive. To
ensure gauge invariance, we must impose $C_A+\tilde{C}_A
=C_{\Pi;0}=0$. For self-energy, both $C_\psi$ and $C_m$ preserve
Lorentz invariance, but are gauge dependent. In our previous
reports, the constants $[C_{\cdots}]$ are termed as "agent"
constants as they should be consequences of taking the "decoupling"
limit of $\{\sigma\}$. To further determine the residual constants,
we must resort to more constraints from symmetries and finally from
physical boundary conditions, which will be pursued in the following
section. Before imposing boundary conditions, the net outcomes of
conventional renormalization programs are also ambiguities, the true
reflection of ill-definedness. The same is also true in EG approach.
However, our strategy leads to the same outcome with {\em neither}
complicated (also unnecessary) manipulation of infinities {\em nor}
modification of Feynman rules at loop levels, hence, it is simple.
So, the real problem is ambiguity (rather than divergence) to be
fixed through appropriate boundary conditions.

Here, some more technical remarks are in order: (1) In the
differential equation approach, any spurious violation of symmetries
due to local ambiguities in the loop amplitudes could be easily
removed. Therefore, the physical breakdown of symmetries (i.e.,
anomalies) must come from {\em definite} properties of the loop
diagrams. Actually, it is shown in our previous studies that chiral
and trace
anomalies\cite{anomaly1,anomaly2,anomaly3,trace1,trace2,trace3,trace4,trace5,trace6}
do originate from the existence of definite or nonlocal terms like
$(k_\mu k_\nu/{k^2})\check{\Gamma}$ in certain divergent
integrals\footnote{Here, $k$ denotes the external momentum at the
corresponding vertex, while $\check{\Gamma}$ the additional
polynomial factors in terms of external momenta at other
vertices.}\cite{Acta951,Acta952,thesis,plb393}. After contraction
with $k^\mu$ or $g^{\mu\nu}$, they give rise to local operators: (a)
$k^\mu \cdot(k_\mu k_\nu/{k^2})\check{\Gamma}=k_\nu\check{\Gamma}$
(chiral anomaly)\cite{Acta951,Acta952,thesis}, (b)
$g^{\mu\nu}\cdot(k_\mu k_\nu/{k^2})\check{\Gamma}=\check{\Gamma}$
(trace anomaly)\cite{plb393,thesis}. In contrast to spontaneous
breaking of global symmetries where massless poles appear at tree
level as consequences, certain massless poles show up in loop
amplitudes as origins of the quantum mechanical violation of
canonical symmetries. (2) Amusingly, this perspective of anomalies
also imply that there may exist properties in divergent integrals
that are both definite (nonlocal) and unexpected from canonical
deductions. Therefore, such divergent diagrams contain more about
quantum field theories\cite{thooft-diagram1,thooft-diagram2}. These
properties might be viewed as indirect but definite links to
underlying theory. (3) The differential equations used here also
imply some structural relations between the Feynman diagrams.
Actually, in QED, as there is no pure gauge boson loops, all the
differentiations with respect to external momenta effectively amount
to inserting the elementary QED vertex, or external photon with zero
momentum. The final result amounts to the Ward identity in
differential form. It will be interesting to see what will happen in
more complicated theories with pure gauge boson loops, like
non-Abelian theories. (4) Technically, one could not obtain a
polynomial/local term from convolution of non-polynomial/nonlocal
functions. A mathematically natural way to yield a polynomial/local
term is to perform certain limit operation. To our interest, this
mathematical scenario corroborates the underlying theory scenario:
The computation should be followed by the "low-energy" limit due to
the wide separation of scales, then some local terms arise from the
"low-energy" limit, indicating the existence of underlying
structures. Conventionally, a regularization is introduced as a
rough substitute of the underlying structures and is taken to zero
in the final stage after removing the "dusts" (infinities) brought
about by the regularization. In our strategy, we simply appreciate
the existence of the underlying structures and then solve the
differential equations that must be satisfied by the limiting
objects.
\subsection{Comparing with Dimensional Regularization} Now we list
out the results from dimensional scheme for a comparison. Our
conclusion is that at one-loop level the only difference between the
two approaches lies in the local part, which is in fact prescription
dependent. The definite (nonlocal) part must be prescription
independent and hence physical. The two approaches are equivalent if
one takes the local part in dimensional scheme as ambiguous. From
this equivalence one could devise a simple method for computation:
do the calculations in dimensional regularization, then replace the
local part with ambiguous polynomials of the corresponding degree.
However, this equivalence is only valid after sub-divergences are
subtracted in case of multi-loop diagrams \cite{ctp38}.

The results from dimensional scheme read, \bea-i\Sigma^{(1)}
(p;\epsilon)=&&\frac{ie^2}{(4\pi)^2}\left\{\int_0^1\!\!\!dz\left[
[(1-2z-\xi)\slash\!\!\!p+(3+\xi)m]\ln\frac{m^2-zp^2}{4\pi\mu^2}
\right.\right.\nonumber\\&&\left.\left.-\frac{(1-\xi)z(p^2-m^2)
\slash\!\!\!p}{m^2-zp^2}\right]+\left(\epsilon^{-1}-\gamma_E
+\frac{1}{2} \right)(\slash\!\!\!p-4m)\right.\nonumber\\&&\left.
-(1-\xi)\left(\epsilon^{-1}-\gamma_E +1\right)(\slash\!\!\!p-m)
\right\}+o(\epsilon),\\
\Lambda^{(1)}_\mu(p,\tilde{p};\epsilon)=&&-\frac{e^2}{(4\pi)^2}
\left\{\int_0^1\!\!\!dy\int_0^{1-y}\!\!\!dz\left[2\gamma_\mu
\ln\Delta+\frac{\mathcal{G}_\mu}{\Delta}+(1-\xi)(1-y-z)\right.
\right.\nonumber\\&&\left.\left.\times\left(-6\gamma_\mu\ln\Delta
+\frac{\mathcal{F}_{1;\mu}}{\Delta^2}+\frac{\mathcal{F}_{2;\mu}}
{\Delta}\right)\right]\right.-\epsilon^{-1}+\gamma_E+2\nonumber\\
&&\left.+(1-\xi)\left(\epsilon^{-1}-\gamma_E-\frac{5}{6}\right)
\gamma_\mu\right\}+o(\epsilon),\\
i\Pi^{(1)}_{\mu\nu}(q;\epsilon)=&&-\frac{8ie^2}{(4\pi)^2}
\left[(q_\mu q_\nu-g_{\mu\nu}q^2)\int^1_0\!\!\!dzz(1-z)\ln
\frac{m^2-z(1-z)q^2}{4\pi\mu^2}\right]\nonumber\\
&&+\frac{4ie^2}{3(4\pi)^2}\left(\epsilon^{-1}-\gamma_E\right) (q_\mu
q_\nu-g_{\mu\nu}q^2)+o(\epsilon).\eea We note that in terms of the
above parametrization, the following correspondence between the
agent constants and the local constants in dimensional
regularization could be easily established:
\bea\label{correspondence}&&\ C_\psi\leftrightarrow
C_\psi(\epsilon)\equiv\xi\left(\epsilon^{-1}-\gamma_E+1
+\ln\left(4\pi\mu^2\right)\right)-\frac{1}{2}, \nonumber\\
&&\frac{C_m}{m}\leftrightarrow\frac{C_m(\epsilon)}{m}\equiv
-(3+\xi)\left(\epsilon^{-1}-\gamma_E+\ln\left(4\pi\mu^2\right)
\right)-1-\xi,\nonumber\\ &&\ C_\perp\leftrightarrow
C_\perp(\epsilon)\equiv-\xi\left(\epsilon^{-1}-\gamma_E
-\frac{5}{6}+\ln\left(4\pi\mu^2\right)\right)+\frac{7}{6},
\nonumber\\&&\ C_A\leftrightarrow C_A(\epsilon)\equiv-\frac{1}{6}
\left(\epsilon^{-1}-\gamma_E+\ln\left(4\pi\mu^2\right)\right).\eea

Obviously, the dimensional regularized results exactly agree with
ours obtained from Eqs.(\ref{diffsf},\ref{diffvt},\ref{diffvp})
after replacing the divergent constants by the agent constants
following the correspondence (\ref{correspondence}). In other words,
the dimensional regularization results (after or before subtraction)
could be seen as a particular solution to our differential equations
at one-loop level. At multi-loop level, the dimensional results will
also satisfy such differential equations loop by loop after all the
corresponding sub-divergences are removed\cite{ctp38}.
\subsection{Reducing ambiguities with Ward identities}
Since gauge invariance is encoded in the Ward identities among
various vertices, it is natural to ask if these identities are
satisfied for the loop amplitudes computed in our approach.
Specifically, we wish to examine the status of the following Ward
identity for self-energy and vertex function,\bea\label{WardId}
\partial_{p^\mu}\Sigma^{(1)}(p)=-\Lambda^{(1)}_\mu(p,p),\eea in our
approach.

To proceed, we make use of the results in dimensional regularization
and look at the Feynman gauge components, i.e., those remain when
$\xi=1$, the rest components will be delegated to Appendix B.
Carrying out the parametric integrations, we have,
\bea-i\Sigma^{(1)}(p;\epsilon)&&=\frac{ie^2}{(4\pi)^2}\left\{\left[
\epsilon^{-1}-\gamma_E+1-\ln\frac{m^2}{4\pi\mu^2}+\delta+(\delta^2-1)
\ln\frac{\delta-1}{\delta}\right]\slash\!\!\!p\right.\nonumber\\
&&\quad\left.-\left[\epsilon^{-1}-\gamma_E+\frac{3}{2}-\ln\frac{m^2}
{4\pi\mu^2}+(\delta-1)\ln\frac{\delta-1}{\delta}\right]4m\right\}
\nonumber\\&&=\frac{ie^2}{(4\pi)^2}\left\{\left[C_\psi(\epsilon)
\|_{\xi=1}+\frac{1}{2}-\ln m^2+\delta+(\delta^2-1)\ln\frac{\delta-1}
{\delta}\right]\slash\!\!\!p\right.\nonumber\\
&&\quad\left.-\left[\frac{C_m(\epsilon)}{4m}\|_{\xi=1}+1-\ln m^2
+(\delta-1)\ln\frac{\delta-1}{\delta}\right]4m\right\},\\
\Lambda^{(1)}_\mu(p,p;\epsilon)&&=\frac{e^2}{(4\pi)^2}\left\{\left[
\epsilon^{-1}-\gamma_E+1-\ln\frac{m^2}{4\pi\mu^2}+\delta +(\delta^2
-1)\ln\frac{\delta-1}{\delta}\right]\gamma_\mu\right.\nonumber\\
&&\quad\left.-2\left[2\left(\delta\ln\frac{\delta-1}{\delta}+1\right)
(\delta\slash\!\!\!p-2m)+\slash\!\!\!p\right]\frac{p_\mu}{p^2}\right\}
\nonumber\\&&=\frac{e^2}{(4\pi)^2}\left\{\left[-C_\perp(\epsilon)
\|_{\xi=1}+3-\ln m^2+\delta +(\delta^2-1)\ln\frac{\delta-1}{\delta}
\right]\gamma_\mu\right.\nonumber\\&&\quad\left.-2\left[2\left(\delta
\ln\frac{\delta-1}{\delta}+1\right)(\delta\slash\!\!\!p-2m)+\slash\!\!\!p
\right]\frac{p_\mu}{p^2}\right\},\eea with $\delta\equiv\frac{m^2}
{p^2}$. The constants $C_\psi(\epsilon), C_\perp(\epsilon)$ have
been defined in the correspondence (\ref{correspondence}). Now
differentiating $\Sigma^{(1)}(p;\epsilon)$ with respect to $p^\mu$,
it is trivial to verify the Ward identity (\ref{WardId}) in
dimensional regularization:\bea-i\partial_{p^\mu}\Sigma^{(1)}
(p;\epsilon)&&= \frac{ie^2}{(4\pi)^2}\left\{\left[\epsilon^{-1}
-\gamma_E+1-\ln \frac{m^2}{4\pi\mu^2}+\delta+(\delta^2-1)
\ln\frac{\delta-1}{\delta}\right]\gamma_\mu\right.\nonumber\\
&&\quad\left.-2\left[2\left(\delta\ln\frac{\delta-1}{\delta}
+1\right)(\delta\slash\!\!\!p-2m)+\slash\!\!\!p\right]
\frac{p_\mu}{p^2}\right\}\nonumber\\&&=i\Lambda^{(1)}_\mu(p,p;
\epsilon).\eea It is easy to see that in dimensional regularization
this Ward identity is ensured by the following relation,
$$C_\psi(\epsilon)\|_{\xi=1}+C_\perp(\epsilon)\|_{\xi=1}
=\frac{5}{2}.$$ In general gauge, we have (Cf. Appendix B):\bea
C_\psi(\epsilon)+C_\perp(\epsilon)=\frac{4+11\xi}{6}.\eea

Evidently, replacing $C_\psi(\epsilon), C_\perp(\epsilon)$ with
$C_\psi,C_\perp$, one enters our approach. Thus the Ward identity
(\ref{WardId}) could hold in our approach only when the agent
constants $C_\psi$ and $C_\perp$ satisfy the same relation as that
between $C_\psi(\epsilon)$ and $C_\perp(\epsilon)$:
\bea\label{C_WardI}C_\psi+C_\perp=\frac{4+11\xi}{6}.\eea That means,
gauge invariance constrains the agent constants $C_\psi$ and
$C_\perp$ according to Eq.(\ref{C_WardI}). Obviously, such
constraints are just welcome in our approach to reduce the
ambiguities. Of course, at higher orders, such constraints must be
also consistently imposed. At one-loop level, there is in principle
no obstacle to remove the local violations of any symmetries. As
removal of the local ambiguities never affects the nonlocal and
hence definite part, any physical anomaly must be originated from
definite and hence prescription-independent sources. As mentioned
above, at least for chiral and trace anomalies, this is indeed the
case: a type of mass-independent and definite terms is the true
source of quantum mechanical violations of the canonical chiral and
scale transformations\cite{Acta951,Acta952,thesis,plb393}.

Therefore, due to Lorentz and gauge invariance, we are left with
three ambiguous constants to fix: $C_\psi$, $C_m$ and $C_A$. This
statement is, however, a little bit cursory before the finiteness of
the rest superficially divergent diagrams, i.e., the three- and
four-photon vertex functions, is established. Below we turn to these
diagrams. Since the three-photon vertex does not contribute to cross
sections or physical observables according to Furry's theorem, we
only need to verify the finiteness of the four-photon vertex, that
is, there is no ultraviolet ambiguity in this vertex function which
is again ensured by gauge invariance.
\subsection{Finiteness of the four-photon vertex}
To verify that the four-photon vertex is free of ambiguity, it
suffices to show that the most divergent piece in each individual
diagram cancel out against each other due to gauge invariance.

The external momenta and polarizations of the four external photon
are arranged in the following manner:
$(p_1,\mu),(p_2,\nu),(p_3,\rho),(p_4,\sigma)$ with
$p_4=-p_1-p_2-p_3$ due to conservation of the four dimensional
momentum, i.e., $p_1$ always enters at vertex $\mu$, $p_2$ at $\nu$,
etc. The four-photon vertex consists of six inequivalent individual
Feynman diagrams that differ from each other by the arrangement of
the relative positions of the external photons along the internal
fermion loop. It is convenient to fix the position of the first
photon of $(p_1,\mu)$ and let the other photons move to obtain the
rest inequivalent loop diagrams. The notations for the four-photon
vertex and the six individual diagrams are as follows,
\bea\Gamma_{\mu\nu\rho\sigma}(p_1,p_2,p_3,p_4)&=&
\Gamma_{1;\mu\nu\rho\sigma}(p_1,p_2,p_3,p_4)
+\Gamma_{2;\mu\rho\nu\sigma}(p_1,p_3,p_2,p_4)
\nonumber\\&&+\Gamma_{3;\mu\nu\sigma\rho}(p_1,p_2,p_4,p_3)
+\Gamma_{4;\mu\sigma\rho\nu}(p_1,p_4,p_3,p_2)\nonumber\\&&
+\Gamma_{5;\mu\sigma\nu\rho}(p_1,p_4,p_2,p_3)
+\Gamma_{6;\mu\rho\sigma\nu}(p_1,p_3,p_4,p_2).\eea The Feynman
integral for $\Gamma_{1;\mu\nu\rho\sigma}$ reads,\bea
\label{4photon1int}i\Gamma_{1;\mu\nu\rho\sigma} =&&(-ie)^4\int
\frac{-d^4l}{(2\pi)^4}\texttt{Tr}\left(\frac{i}{\slash\!\!l-m
+i\epsilon}\gamma_\mu\frac{i}{\slash\!\!l+\slash\!\!\!p_1-m
+i\epsilon}\gamma_\nu\right.\nonumber\\&&\times\left.
\frac{i}{\slash\!\!l+\slash\!\!\!p_1+\slash\!\!\!p_2-m+i\epsilon}
\gamma_\rho\frac{i}{\slash\!\!l+\slash\!\!\!p_1+\slash\!\!\!p_2
+\slash\!\!\!p_3-m+i\epsilon}\gamma_\sigma\right).\eea By
rearranging the external momenta and the elementary vertices, one
could readily obtain the integrals for the rest five diagrams. Since
the superficial divergence degree of such diagrams is 0, then it
suffices to verify that the logarithmic divergence is absent in the
sum of the six diagrams in order that gauge invariance is preserved.
Obviously, it suffices to do computation with
$\Gamma_{1;\mu\nu\rho\sigma}$ as others could be obtained by
permutations of the vertices $(p_1,\mu),(p_2,\nu),\cdots$.

Differentiating the integral in Eq.(\ref{4photon1int}) with respect
to $p_{1;\alpha}$, the resulting integral becomes definite and
gives,\bea\partial_{p_{1;\alpha}}i\Gamma_{1;\mu\nu\rho\sigma}&&=
\frac{8ie^4}{(4\pi)^2}T_{1;\mu\nu\rho\sigma}\int_0^1\!\!\!dx
\int_0^x\!\!\!dy\int_0^y\!\!\!dz\partial_{p_{1;\alpha}}\ln\tilde
{\Delta}_1+\partial_{p_{1;\alpha}}i\check{\Gamma}_{1;\mu\nu\rho
\sigma},\\T_{1;\mu\nu\rho\sigma}&&\equiv g_{\mu\nu}g_{\rho\sigma}
-2g_{\mu\rho}g_{\nu\sigma}+g_{\mu\sigma}g_{\nu\rho},\\
\tilde{\Delta}_1&&\equiv m^2-(x-x^2)p_1^2-(y-y^2)p_2^2-(z-z^2)p_3^2
\nonumber\\&&\quad-2y(1-x)p_1p_2-2z(1-x)p_1p_3-2z(1-y)p_2p_3,\eea
where the $\check{\Gamma}_{1;\mu\nu\rho\sigma}$ denotes the definite
parts after integrating back with respect to $p_{1;\alpha}$ and is
not concerned here. Next, integrating back with respect to
$p_{1;\alpha}$, the diagram $\Gamma_{1;\mu\nu\rho\sigma}$ could be
parametrized as follows:\bea i\Gamma_{1;\mu\nu\rho\sigma}
=\frac{8ie^4}{(4\pi)^2}T_{1;\mu\nu\rho\sigma}\int_0^1\!\!\!dx
\int_0^x\!\!\!dy\int_0^y\!\!\!dz\ln\frac{\tilde{\Delta}_1}
{C_{1;4\gamma}}+\check{\Gamma}_{1;\mu\nu\rho\sigma},\eea where an
integration constant $C_{1;4\gamma}$ is shown explicitly, which is
exactly the ambiguity that corresponds to the logarithmic divergence
of the integral in Eq.(\ref{4photon1int}). The results for other
diagrams are given in Appendix C. For later convenience, we
integrate by parts with respect to the Feynman parametric
integration to arrive at the following expressions for
$\Gamma_{1;\mu\nu\rho\sigma}$, \bea\label{4gamma-repres}
i\Gamma_{1;\mu\nu\rho\sigma}=\frac{4ie^4} {3(4\pi)^2}
T_{1;\mu\nu\rho\sigma} \ln\frac{m^2}{C_{1;4\gamma}}
+\tilde{\Gamma}_{1;\mu\nu\rho\sigma},\eea where the definite terms
arising from this operation have been absorbed into
$\check{\Gamma}_{1;\mu\nu\rho\sigma}$ and collectively denoted as
$\tilde{\Gamma}_{1;\mu\nu\rho\sigma}$, cf. Appendix D.

Now, in order that the ill-defined piece in the full amplitude
vanish\bea i\Gamma_{\mu\nu\rho\sigma}\|_{\texttt{ill-defined}}&&
=\frac{4ie^4}{3(4\pi)^2}\left[\sum_{k=1}^6T_{k;\mu\nu\rho\sigma}
\ln\frac{m^2}{C_{k;4\gamma}}\right]=0,\eea with $\sum_{k=1}^6
T_{k;\mu\nu\rho\sigma}=0$, the constant $C_{k;4\gamma}$ must be
identical for each $k$ (say, $C_{k;4\gamma}=\bar{C}_{4\gamma},
\forall k$). Otherwise, gauge invariance will be violated. Thus,
similar as the case of vacuum polarization for two-photon vertex,
gauge invariance also remove the ambiguities in the four-photon
vertex function.

In fact, after setting all the external momenta to zero, we will end
up with the same ill-defined integral in all the six diagrams:
\bea\int\frac{d^4l}{(2\pi)^4}\frac{l_al_bl_cl_d}{(l^2-m^2)^4}=-
\frac{i}{24(4\pi)^2}[g_{ab}g_{cd}+\cdots]\ln\frac{m^2}
{\bar{C}_{4\gamma}}.\eea Thus it is natural to deem that the
constants $C_{k;4\gamma}$ are identical in any reasonable
prescription.

In dimensional regularization, one would find the following
correspondence in each of the six diagrams:
$$\ln\frac{m^2}{C_{k;4\gamma}}\leftrightarrow-\epsilon^{-1}
+\gamma_E+\ln\frac{m^2}{{4\pi\mu^2}},$$ which means the full
four-photon vertex is finite in dimensional regularization due to
the mechanism demonstrated above.
\subsection{Brief summary of differential equation approach to QED
at one-loop level} Now, we have treated all the superficially
divergent or ill-defined loop diagrams of QED at one-loop level,
without resorting to any artificial deformation of spacetime and
symmetries or any form of cutoffs. At one-loop level, only three
local ambiguities remain to be fixed in QED: $C_\psi$, $C_m$ and
$C_{A}$ that appear in self-energy $\Sigma$, vertex $\Lambda_\mu$
and vacuum polarization $\Pi_{\mu\nu}$, respectively. This is a {\em
natural} conclusion following from Lorentz and gauge invariance and
the existence of a complete underlying theory.

In particular, from our treatment, one could naturally infer that,
any fundamental theory that is well defined and underlies the
presently known QED must also yield the same or equivalent
functional expressions as given here after taking the "decoupling"
limit, the only difference is that the unknown agent constants could
be unambiguously computed in such underlying theory. Therefore, in
the underlying theory perspective, the QED processes defined by the
elementary Feynman diagrams (in covariant gauge) represented by
Fig.1 must take the following form, \bea-i\Sigma^{(1)}(p)=&&\frac{
ie^2}{(4\pi)^2}\left\{\int_0^1\!\!\!dz\left[[(1-2z-\xi)\slash
\!\!\!p+(3+\xi)m]\ln\frac{m^2-zp^2}{\mu^2_{\text{UD}}}\right.\right.
\nonumber\\&&\left.\left.-\frac{(1-\xi)z(p^2-m^2)\slash\!\!\!p}
{m^2-zp^2}\right]+c^0_\psi\slash\!\!\!p+c^0_mm\right\},\\
\Lambda^{(1)}_\mu(p,\tilde{p})=&&-\frac{e^2}{(4\pi)^2}\left\{
\int_0^1\!\!\!dy\int_0^{1-y}\!\!\!dz\left[2\gamma_\mu
\ln\Delta+\frac{\mathcal{G}_\mu}{\Delta}+(1-\xi)(1-y-z)\right.
\right.\nonumber\\&&\left.\left.\times\left(-6\gamma_\mu\ln\Delta
+\frac{\mathcal{F}_{1;\mu}}{\Delta^2}+\frac{\mathcal{F}_{2;\mu}}
{\Delta}\right)\right]+\left(\frac{4+11\xi}{6}
-c^0_\psi\right)\gamma_\mu\right\},\\
i\Pi^{(1)}_{\mu\nu}(q)=&&\frac{8ie^2(g_{\mu\nu}q^2-q_\mu q_\nu)
}{(4\pi)^2}\left[\int^1_0\!\!\!dzz(1-z)\ln\frac{m^2-z(1-z)q^2}
{\mu^2_{\text{UD}}}+c^0_A\right],\nonumber\\\eea where we have
explicitly "separated out" a dimensional constant $\mu_{\text{UD}}$
from all the constants $C_{\cdots}$ to balance the dimension in the
logarithmic parts, the residual parts are purely dimensionless
numbers, each denoted as $c^0_{\cdots}$. This way of parametrizing
the constants will be more precise and useful for the following
discussions.

We should also note that, in the underlying theory perspective, all
the constants or parameters that appear in the Lagrangian should be
defined from the "decoupling" limit of the underlying theory also as
"agents". In our view, these "agents" in lagrangian are only
elementary from the perspective of (effective) field theories, not
necessary so in the perspective of underlying theory. As the agent
constants appear in the local part of the elementary vertex
functions within perturbative framework, the Lagrangian constants
would mix with these local terms containing the agent constants in
the corresponding vertex functions. This is in fact the origin of
renormalization group equations in our approach, without resorting
to the conventional renormalization theory, see, Sec. IV for a brief
discussion.
\section{Removing ambiguities and "renormalization" scheme}
Now we turn to determining the ambiguous agent constants. Obviously,
the conventional renormalization schemes would be reproduced
provided that the ambiguous constants as well as mass and couplings
are fixed through the conventional renormalization conditions. An
appropriate choice of the whole set $\{m,e;\mu,c^0_{\cdots}\}$
defines a renormalization prescription or scheme. So, our simple
strategy is compatible with the conventional renormalization
programs. In principle, any scheme can be employed provided the same
scattering matrix elements or the same physical observables could be
obtained. This freedom just leads to the renormalization group in
the general sense of St\"uckelberg-Peterman\cite{SP}. Fixing the
dimensionless constants in a scheme and letting the scale
$\mu_{\text{UD}}$ vary, one could obtain the renormalization group
equations in the narrow sense.

To proceed, the following parametrization would be used useful,\bea
\Sigma^{(1)}(p)&&=A^{(1)}(p^2)\slash\!\!\!p+B^{(1)}(p^2)m,\\
\bar{u}(p)\Lambda^{(1)}_\mu u(\tilde{p})&&=\bar{u}(p)\gamma_\mu
u(\tilde{p})F^{(1)}(k^2)+\frac{i\bar{u}(p)\sigma_{\mu\nu}k^\nu
u(\tilde{p})}{2m}G^{(1)}(k^2),\\
\Pi^{(1)}_{\mu\nu}(q)&&=-\Pi^{(1)}(q^2)(g_{\mu\nu}q^2-q_\mu q_\nu),\\
A^{(1)}(p^2)&&\equiv-\frac{\alpha}{4\pi}\left[\xi\left(\left(
\delta^2-1\right)\ln\frac{\delta-1}{\delta}+\delta-\ln\frac{m^2}
{\mu^2_{\text{UD}}}\right)+\frac{1}{2}+c^0_\psi\right],\\
B^{(1)}(p^2)&&\equiv\frac{\alpha}{4\pi}\left[(3+\xi)\left((\delta-1)
\ln\frac{\delta-1}{\delta}+1-\ln\frac{m^2}{\mu^2_{\text{UD}}}\right)
-c^0_m\right],\\
F^{(1)}(k^2)&&\equiv\frac{\alpha}{4\pi}\left[2\text{arccoth}[\theta_k]
\left[\left(1+\ln\frac{\lambda_\gamma}{m^2}\right)\left(\theta_k
+\frac{1}{\theta_k}\right)+\theta_k\right]+2\right.\nonumber\\
&&\left.\quad-{\Xi}(k^2;1-\xi)-\ln\frac{m^2}{\mu^2_{\text{UD}}}
-c^0_\perp\right],\\
G^{(1)}(k^2)&&\equiv\frac{\alpha}{2\pi}\left(\theta_k-\frac{1}
{\theta_k}\right)\text{arccoth}[\theta_k],\\
\Pi^{(1)}(q^2)&&\equiv\frac{\alpha}{3\pi}\left[\theta_q\left(
\theta_q^2-3\right)\text{arccoth}[\theta_q]-\theta_q^2-\ln\frac{m^2}
{\mu^2_{\text{UD}}}+\frac{8}{3}-6c^0_A\right],\\
\delta&&=\frac{m^2}{p^2},\ k=p-\tilde{p},\ \theta_x\equiv
\sqrt{1-\frac{4m^2}{x^2}}.\eea Note that the vertex function is
sandwiched between on-shell states $[u(\tilde{p}),\bar{u}(p)]$ of
electron. Obviously, the scalar factors $A^{(1)}, B^{(1)}, F^{(1)}$
and $\Pi^{(1)}$ contain ultra-violet ambiguities (depending on
$c^0_\psi, c^0_m, c^0_\perp$ and $c^0_A$, respectively). Moreover,
the first three factors are also gauge dependent and infrared
singular when the external momenta are on mass shell. In $F^{(1)}$,
we have explicitly put in a photon mass squared ($\lambda_\gamma$)
to regulate the infrared singularity, with ${\Xi}$ denoting the
terms proportional to $1-\xi$. In contrast, the scalar factor
$G^{(1)}$ is both ultra-violet and infrared finite and also gauge
independent. Its value at zero momentum transfer ($k^2=0$) gives the
well-known anomalous magnetic moment of electron at one-loop level:
$$G^{(1)}(0)=\frac{\alpha}{2\pi}.$$
\subsection{Various boundary conditions for vertex function}
Now we impose various conditions on the elementary vertices (here,
$\Sigma^{(1)}, \Lambda^{(10}_\mu, \Pi^{(1)}_{\mu\nu}$)\footnote{They
could be found in many textbooks for field theory, see, e.g.
Ref.\cite{peskin}}.  In doing so, both the tree level parameters
($m, e$) and the agent constants must be understood as
scheme-dependent so that the observables computed using these
parameters and vertices are physical and independent of schemes. The
lagrangian or tree level mass or coupling needs not be just the
physical one, which may differ by a "(re)normalization" constant in
the conventional terminology. In our strategy, such
"(re)normalization" constants could also be introduced but are
nevertheless ultra-violet finite. Actually, as is already pointed
out in Ref.\cite{weinbergvol1}: "the renormalization of mass and
field has nothing directly to do with the presence of infinities,
and would be necessary even in a theory in which all momentum space
integrals are convergent".
\subsubsection{On-shell conditions} First, let us consider the
so-called on-shell conditions in Feynman gauge, which read,\bea
\Sigma^{(1)}\|_{\slash\!\!\!p=m}=0,\ \partial_{\slash\!\!\!p}
\Sigma^{(1)}\|_{\slash\!\!\!p=m}=0,\ \Lambda^{(1)}_\mu
\|_{\slash\!\!\!p=\slash\!\!\!\tilde{p}=m}=0,\ \Pi^{(1)}
\|_{q^2=0}=0.\eea Using the expressions given above, we could find
that:\bea\Sigma^{(1)}\|_{\slash\!\!\!p=m}&&=\frac{\alpha}{4\pi}
\left(-3\ln\frac{m^2}{\mu^2_{\text{UD}}}+\frac{5}{2}-c^0_\psi
-c^0_m\right)m,\nonumber\\ \partial_{\slash\!\!\!p}\Sigma^{(1)}
\|_{\slash\!\!\!p=m}&&=\frac{\alpha}{4\pi}\left(\ln\frac{m^2}
{\mu^2_{\text{UD}}}-2\ln\frac{\lambda_\gamma}{m^2}-\frac{7}{2}
-c^0_\psi\right),\nonumber\\ \Lambda^{(1)}_\mu\|_{\slash\!\!\!p=
\slash\!\!\!\tilde{p}=m}&&=\frac{\alpha}{4\pi}\left(-\ln\frac{m^2}
{\mu^2_{\text{UD}}}+2\ln\frac{\lambda_\gamma}{m^2}+\frac{7}{2}
+c^0_\psi\right)\gamma_\mu,\nonumber\\ \Pi^{(1)}\|_{q^2=0}&&
=\frac{\alpha}{3\pi}\left(-\ln\frac{m^2}{\mu^2_{\text{UD}}}
-6c^0_A\right),\eea where $c^0_\psi+c^0_\perp=5/2$ has been used.
Note that all the infrared singularities have been regularized with
a photon mass, and $m$ now denotes the physical mass for electron.
According to conventional programs, the coupling $e$ or
$\alpha=e^2/(4\pi)$ appearing in these conditions is understood to
be the physical one.

Then the on-shell conditions could be fulfilled provided the
constants $[c^0_{\cdots}]$ take the following values:\bea&&
c^0_\psi=\ln\frac{m^2}{\mu^2_{\text{UD}}}-2\ln\frac{\lambda_\gamma}
{m^2}-\frac{7}{2}=\frac{5}{2}-c^0_\perp,\nonumber\\
&&c^0_m=-4\ln\frac{m^2}{\mu^2_{\text{UD}}}+2\ln\frac{
\lambda_\gamma}{m^2}+6,\quad c^0_A=-\frac{1}{6}\ln
\frac{m^2}{\mu^2_{\text{UD}}}.\eea In this manner, all the
ambiguities are gone in the vertices, including the scale
$\mu_{\text{UD}}$:\bea A^{(1)}_{os}(p^2)&&\equiv-\frac{\alpha}
{4\pi}\left[\left(\delta^2-1\right)\ln\frac{\delta-1}
{\delta}+\delta-3-2\ln\frac{\lambda_\gamma}{m^2}\right],\\
B^{(1)}_{os}(p^2)&&\equiv\frac{\alpha}{4\pi}\left[4(\delta-1)
\ln\frac{\delta-1}{\delta}-2-2\ln\frac{\lambda_\gamma}{m^2}\right],
\\ F^{(1)}_{os}(k^2)&&\equiv\frac{\alpha}{4\pi}\left[2\text{arccoth}
[\theta_k]\left[\left(1+\ln\frac{\lambda_\gamma}{m^2}\right)\left(
\theta_k +\frac{1}{\theta_k}\right)+\theta_k\right]\right.
\nonumber\\&&\quad\left.-4-2\ln\frac{\lambda_\gamma}{m^2}\right]\\
\Pi^{(1)}_{os}(q^2)&&\equiv\frac{\alpha}{3\pi}\left[\theta_q\left(
\theta_q^2-3\right)\text{arccoth}[\theta_q]-\theta_q^2+\frac{8}{3}
\right].\eea The on-shell scheme works for field theories without
unstable fields as elementary quantum degrees of freedom. The tree
level mass is identified as the physical pole mass. But for
complicated theories like electroweak with unstable sectors, the
on-shell scheme runs into trouble, see
Ref.\cite{unstable1,unstable2,unstable3,unstable4,unstable5,unstable6,unstable7,unstable8,unstable9,unstable10}.
\subsubsection{"Minimal" conditions}
In analogy to the famous "Minimal Subtraction", one may define a
scheme where $[c^0_{\cdots}]$ are removed so that only the
$\sim\ln\mu^2$ pieces are left over, which will lead to the
well-known mass-independent running behavior. We will denote this
scheme as "Ms", which may also be interpreted as "Minimal
specification".

Since $c^0_\psi$ and $c^0_\perp$ can not be zero at the same time
due to gauge invariance, we choose the following:\bea c^0_\psi
+\frac{1}{2}=c^0_m=c^0_A=0,\quad c^0_\perp=3.\eea Then the
prescription-dependent scalar factors would take the following
neater forms (in Feynman gauge):\bea A^{(1)}_{Ms}(p^2)&&\equiv
-\frac{\alpha}{4\pi}\left[\left(\delta^2-1\right)\ln\frac{\delta-1}
{\delta}+\delta-\ln\frac{m^2}{\mu^2_{Ms}}\right],\\
B^{(1)}_{Ms}(p^2)&&\equiv\frac{\alpha}{\pi}\left[(\delta-1)
\ln\frac{\delta-1}{\delta}+1-\ln\frac{m^2}{\mu^2_{Ms}}\right],\\
F^{(1)}_{Ms}(k^2)&&\equiv\frac{\alpha}{4\pi}\left[2\text{arccoth}
[\theta_k]\left[\left(1+\ln\frac{\lambda_\gamma}{m^2}\right)
\left(\theta_k +\frac{1}{\theta_k}\right)+\theta_k\right]\right.
\nonumber\\&&\quad\left.-1-\ln\frac{m^2}{\mu^2_{Ms}}\right],\\
\Pi^{(1)}_{Ms}(q^2)&&\equiv\frac{\alpha}{3\pi}\left[\theta_q\left(
\theta_q^2-3\right)\text{arccoth}[\theta_q]-\theta_q^2+\frac{8}{3}
-\ln\frac{m^2}{\mu^2_{Ms}}\right].\eea Here the subscript "$Ms$" for
mass $m$ and coupling $e$ are omitted to avoid heavy symbolism and
the scale $\mu_{\text{UD}}$ is now replaced by $\mu_{Ms}$. Of
course, the concrete values for $m$ and $e$ in this scheme need to
be determined in terms of physical values of mass using appropriate
physical observables\cite{Ster}. In this prescription, the scalar
factors $A$ and $B$ at general off-shell momenta are not beset with
infrared singularities in contrast to the on-shell prescription.
\subsubsection{Boundary conditions at vanishing off-shell momenta}
One could also impose boundary conditions at off-shell momenta.
Among these off-shell prescriptions, the zero momenta case is much
simpler for QED with massive fermion. In Feynman gauge, we have:
\bea\Sigma^{(1)}\|_{\slash\!\!\!p=0}&&=\frac{\alpha}{4\pi}
\left(-4\ln\frac{m^2}{\mu^2_{\text{UD}}}-c^0_m\right)m,\nonumber\\
\partial_{\slash\!\!\!p}\Sigma^{(1)}\|_{\slash\!\!\!p=0}&&
=\frac{\alpha}{4\pi}\left(\ln\frac{m^2}{\mu^2_{\text{UD}}}
-c^0_\psi\right),\nonumber\\
\Lambda^{(1)}_\mu\|_{\slash\!\!\!p=\slash\!\!\!\tilde{p}=0}
&&=\frac{\alpha}{4\pi}\left(-\ln\frac{m^2}{\mu^2_{\text{UD}}}
+\frac{5}{2}-c^0_\perp\right)\gamma_\mu,\nonumber\\
\Pi^{(1)}\|_{q=0}&&=\frac{\alpha}{3\pi}\left(-\ln\frac{m^2}
{\mu^2_{\text{UD}}}-6c^0_A\right),\eea so they vanish if \bea
c^0_\psi=\ln\frac{m^2}{\mu^2_{\text{UD}}}=\frac{5}{2}-c^0_\perp,\
c^0_m= -4\ln\frac{m^2}{\mu^2_{\text{UD}}},\ c^0_A=-
\frac{1}{6}\ln\frac{m^2}{\mu^2_{\text{UD}}}.\eea Then the scalar
factors read:\bea A^{(1)}_{zm}(p^2)&&\equiv-\frac{\alpha}{4\pi}
\left[\left(\delta^2-1\right)\ln\frac{\delta-1}{\delta}+\delta
+\frac{1}{2}\right],\\
B^{(1)}_{zm}(p^2)&&\equiv\frac{\alpha}{\pi}\left[(\delta-1)
\ln\frac{\delta-1}{\delta}+1\right],\\
F^{(1)}_{zm}(k^2)&&\equiv\frac{\alpha}{4\pi}\left[2\text{arccoth}
[\theta_k]\left[\left(1+\ln\frac{\lambda_\gamma}{m^2}\right)
\left(\theta_k +\frac{1}{\theta_k}\right)+\theta_k\right]
-\frac{1}{2}\right],\\
\Pi^{(1)}_{zm}(q^2)&&\equiv\frac{\alpha}{3\pi}\left[\theta_q\left(
\theta_q^2-3\right)\text{arccoth}[\theta_q]-\theta_q^2
+\frac{8}{3}\right].\eea The nice point of off-shell conditions at
zero momenta (hence the subscript "$zm$") is that, the infrared
singularities do not show up in the scalar factors at general
off-shell momenta, just like in the "$Ms$" prescription. Such
conditions possess an intuitive interpretation: at zero momenta,
quantum fluctuations are suppressed, and an electron would serve as
a classical static source of electromagnetic field, with the tree
level coupling "$e$" being pertinent to static or Coulomb charge
somehow.
\subsubsection{Physical boundary conditions in terms of observables}
In principle, boundary conditions may be imposed upon many other
field-theoretical objects that are constructed with elementary
vertex functions or Green function. It would be better to work with
the observables that could be readily confronted with physical data
and depend on the elementary parameters (masses, couplings, and
agent constants) in an as simple as possible manner, sparing the
intermediate renormalization.

For example, one may proceed as below in QED: First, physical values
are assigned to the Lagrangian parameters (masses and couplings),
the agent constants are treated as unknown or arbitrary; Second,
appropriate observables are selected that could be readily expressed
in terms of Feynman diagrams that definitely contain the agent
constants; Finally, the agent constants could be fixed in terms of
the physical parameters and data of these observables. In
formulae,\bea(1)&&\quad\{m,e;
[C_{\cdots}]\}\Rightarrow\{m^{\text{(phy)}},e^{(\text{phy})};
[C^{(\text{phy})}_{\cdots}]\},\nonumber\\(2)&&\quad Q_n\left(
m^{(\text{phy})},e^{(\text{phy})};[C^{(\text{phy})}_{\cdots}]
\right)=Q^{\texttt{(data)}}_n, \ n=1,2, \cdots\nonumber\\
&&\Rightarrow C^{(\text{phy})}_{\cdots}=f_{\cdots}\left(
[Q^{\texttt{(data)}}_n]; m^{(\text{phy})},e^{(\text{phy})}
\right).\eea The set of observables to be employed for this purpose
should fulfill some natural requirements such as simplicity and
analyticity with respect to their dependence upon mass, coupling and
the agent constants, well-definedness in the infrared, etc. In
principle, even theories like quantum chromodynamics (QCD) that are
elusive of physical pole masses of elementary quantum fields can be
treated in this manner, where the physical contents of the
lagrangian or tree level masses and couplings becomes a nontrivial
issue. Such imposition of physical boundary conditions is general
and natural from underlying theory perspective. Actually, this is
exactly what is done in literature for the determination of the
"physical" coupling of strong interaction, see, e.g.,
Ref.\cite{Ster}, chapter 12.
\subsubsection{A remark on boundary conditions}
In order that the same observables are obtained, different choices
of "intermediate" boundary conditions or prescriptions for vertex
functions must be related to each other somehow, or more
specifically, the transformations across different prescriptions
defined by the boundary conditions should in principle be feasible,
at least within perturbation theory. In conventional programs, this
is the so-called scheme dependence in perturbation theory, where
different renormalization conditions or prescriptions, differ by a
set of finite renormalization constants, and the observables
computed in different prescriptions should agree with each other to
the perturbation order computed. This is in fact encoded in the
renormalization group equation in the sense of
St\"uckelberg-Peterman as mentioned in the very beginning of this
section. Furthermore, a general prescription would also contain an
arbitrary scale as "renormalization point". The variation of this
arbitrary scale within the specified prescription should not affect
observables, which would in effect lead to renormalization group
equations that are themselves prescription- or scheme-dependent.

The above statements make perfect sense within perturbative
framework. However, things might become complicated in
nonperturbative contexts. As an simple example, we refer to our
previous work\cite{PRD65}, where it is shown that some scheme in
incompatible with dynamical symmetry breaking of massless
$\lambda\phi^4$ theory in nonperturbative context. Strictly
speaking, any intermediate boundary conditions or prescriptions to
be employed must be in the same "orbit" as the physical boundaries.
This in turn requires a complete solution of the full theory of
quantum fields, which is presently out of our reach. So, scheme
dependence is a real issue for quantum field theories beyond
perturbative regime.
\subsubsection{Influences of infrared divergences} It is well-known
that QED is beset with infrared divergences. For on-shell scheme,
infrared divergences show up already in the boundary conditions for
the loop amplitudes, as is evident above. Simply speaking, infrared
singularity arises because Fock states could not work well in
presence of massless particles. Thus introducing certain extent of
coherence, the infrared divergences in QED could be removed by use
of Bloch-Nordsieck theorem\cite{BN}, which has been described in
detail in many standard field theory textbooks, e.g.,
\cite{Ster,peskin,weinbergvol1}. At one loop level, one regularize
the infrared singularities in the loop diagrams somehow (here in
this report, in terms of a fictitious photon mass) and they will
cancel out against those from real diagrams in physical observables.

Of course, for more complicated non-Abelian gauge theories, severe
infrared divergences are present and their treatments are more
involved. Nevertheless, as our approach do not alter anything in the
infrared, so all the conventional methods for dealing with infrared
singularities or similar singularities except the ultraviolet ones
may well apply. As is clear from our above discussions, one may work
with a set of renormalization conditions that are not plagued by
infrared singularities to avoid further complications, this could be
done with conditions or "boundaries" defined at off-shell momenta
where everything is infrared finite\cite{KPQ1,KPQ2}, e.g., at zero
four-momenta as illustrated above.
\subsection{Lagrangian perspective}
In this subsection, we reexamine the issue in terms of lagrangian or
action, which allows us to examine a number of issues in a more
transparent manner. Such effort is also helpful to resolve some
conceptual difficulties that have been "afflicting" the conventional
practices.
\subsubsection{Brief review of conventional programs}
The conventional program of renormalization starts with a bare
lagrangian:\bea\mathcal{L}(\psi_B,A_B^\mu;m_B,e_B)={\bar\psi}_B(
i\slash\!\!\!\!D_B -m_B)\psi_B -\frac{1}{4}F_{B;\mu\nu}F_B^{\mu\nu}
-\frac{1}{2\xi_B}(\partial_\mu A_B^\mu)^2, \eea with
$D^\mu_B=\partial^\mu-ie_B A_B^\mu$. Then all the 1PI vertices
$\left(\Gamma^n_B([p_1,\cdots,p_n],m_B,e_B)\right)$ are renormalized
through the following renormalization rescaling,\bea\psi_B
=Z_\psi^{1/2}\psi,\ A_B^\mu=Z_A^{1/2}A^\mu,\ m_B=\frac{Z_m}
{Z_\psi}m,\ e_B=(Z_A)^{-1/2}e,\ \xi_B=Z_A\xi,\eea and\bea
\mathcal{L}(\psi_B,A_B^\mu;m_B,e_B)&&=Z_\psi{\bar\psi}i\slash\!\!\!\!D
\psi-Z_mm{\bar\psi}\psi-\frac{Z_A}{4}F_{\mu\nu}F^{\mu\nu}-\frac{1}
{2\xi}(\partial_\mu A^\mu)^2\nonumber\\&&=\mathcal{L}(\psi,A^\mu;
m,e)+\Delta\mathcal{L}(\psi,A^\mu;Z_\psi-1,Z_m-1,Z_A-1),\eea so that
\bea\Gamma^{(n)}_B([p_1,\cdots,p_n],m_B,e_B)=Z^{-n_\psi/2}_\psi
Z^{-n_A/2}_A\Gamma^{(n)}([p_1,\cdots,p_n],m,e).\eea Here,
$\Delta\mathcal{L}$ serves to provide all the necessary
counter-terms for canceling out any divergences in the loop
diagrams. As the BPHZ algorithm is equivalent to the above program,
below, we will refer to both as conventional approaches or programs.
One caveat is that the formalism for such programs is only
established within the realm of perturbation and hence must be so
implemented.
\subsubsection{Underlying theory perspective}Now, let us reexamine
the same issue from underlying theory perspective. It is convenient
to work with the path integral formulation. In terms of a underlying
theory description, the generating functional for the QED processes
would formally take the following form:\bea Z(J_\mu,\eta,\bar{
\eta};\{\sigma\})=&&\int\!\!d\mu[\phi_{\text{UT}}]\exp\left\{i\int
\!\!d^4x\left[\mathcal{L}_{\text{UT}}([\phi_{\text{UT}}],
\{\sigma\})\right.\right.\nonumber\\&&\left.\left.+J_\mu
O^\mu_{A}(\{\sigma\})+O_{\bar{\psi}}(\{\sigma\})
\eta+\bar{\eta}O_{\psi}(\{\sigma\})\right]\right\},\eea where
$[\phi_{\text{UT}}]$ is used to label all the necessary degrees of
freedom, and $O^\mu_A, O_{\bar{\psi}}, O_\psi$ denotes the photon
and electron field parameters in the UT description which may well
be composite ones.

At "lower" scales where QED become prominent or effective processes,
all the typical higher energy modes are actually integrated out,
resulting in a well-defined functional in terms of dominant or
effective degrees,\bea Z(J_\mu,\eta,\bar{\eta};\{\sigma\})=
&&\int\!\! dA^\mu d\bar{\psi}d\psi\exp\left\{i\int\!\!d^4x\left[
\mathcal{L}_{\text{eff}}(\psi,A_\mu;\{\sigma\})\right.\right.
\nonumber\\&& \left.\left.+J_\mu A^\mu+\bar{\psi}\eta
+\bar{\eta}\psi\right]\right\},\eea with $\int\!d^4x
\mathcal{L_{\text{eff}}}(\psi,A_\mu;\{\sigma\})$ being the effective
action generated from the integrating-out. Taking the "decoupling
limit" ($\breve{\mathcal{P}}_{\texttt{\scriptsize LE}}$) na\"{i}vely
on the effective lagrangian would yield to a classical lagrangian of
QED,\bea\breve{\mathcal{P}}_{\texttt{\scriptsize LE}}[\mathcal{L}
_{\text{eff}}(\psi,A_\mu;\{\sigma\})]=\mathcal{L}_{\text{QED}}
(\psi,A_\mu),\eea which in turn leads to ill-defined path integral
as the limit operation and the functional integration does not
commute. Obviously, the details taken away with the limit operation,
\bea\delta\mathcal{L}(\psi,A_\mu;\{\sigma\})\equiv&&\mathcal{L}
_{\text{eff}}(\psi,A_\mu;\{\sigma\})-\breve{\mathcal{P}}
_{\texttt{\scriptsize LE}}[\mathcal{L}_{\text{eff}}(\psi,A_\mu;
\{\sigma\})]\nonumber\\=&&\mathcal{L}_{\text{eff}}(\psi,A_\mu;\{\sigma\})
-\mathcal{L}_{\text{QED}}(\psi,A_\mu),\nonumber\eea is just what we
are missing for a well defined path integral. Therefore, we must
include it into the path integral, \bea\label{Z-delta-UT}
Z(J_\mu,\eta,\bar{\eta};\{\sigma\})=&&\int\!\!dA^\mu d\bar{\psi}
d\psi\exp\left\{i\int\!\!d^4x\left[\mathcal{L}_{\text{QED}}(\psi,
A_\mu)+\delta\mathcal{L}(\psi,A_\mu;\{\sigma\})\right.\right.
\nonumber\\&&\left.\left.+J_\mu A^\mu+\bar{\psi}\eta
+\bar{\eta}\psi\right] \right\}.\eea Evidently, in conventional
programs, $\delta \mathcal{L}$ is implemented through regularization
and subtraction that are encoded in $\Delta\mathcal{L}$. For certain
interactions, the job of $\Delta\mathcal{L}$ could be "autonomously"
done using operators appearing in the classical lagrangian, leading
in effect to the "renormalization" of these operators. Such
autonomous cases are conventionally termed as renormalizable, while
the rest as un-renormalizable. As a "realization" or substitute of
$\delta\mathcal{L}$, $\Delta\mathcal{L}$ must also take the
responsibility to clear all the {\em side effects} like ultraviolet
infinities associated with a regularization. Only in this sense, the
manipulation of infinities (through counter terms) may be
justifiable, and the bare lagrangian may serve as a symbolic and
rough representation of the underlying theory:
$\mathcal{L}_{\text{eff}}(\psi,A_\mu;\{\sigma\})$, which definitely
contains the sophisticated underlying {\em quantum} details that
could not be simply described in terms of "bare" objects that are
originally coined in {\em classical} field theory.

As we could at best obtain the parametric form of the "decoupling"
effects of $\{\sigma\}$, there may be certain arbitrariness in our
choice of $\mathcal{L}_{\text{QED}}=\breve{\mathcal{P}}
_{\texttt{\scriptsize LE}}[\mathcal{L}_{\text{eff}}]$ (that is
actually unknown to us yet) as our "starting" lagrangian for
calculations or the in the separation of it out of $\mathcal{L}
_{\text{eff}}$:$$\mathcal{L}_{\text{eff}}=\tilde{\mathcal{L}}
_{\text{QED}}+\tilde{\delta}\mathcal{L}(\cdots;\{\sigma\})
=\mathcal{L}_{\text{QED}}+\delta \mathcal{L}(\cdots;\{\sigma\}),$$
where $\tilde{\mathcal{L}}_{\text{QED}}$ and $\mathcal{L}
_{\text{QED}}$ would at most differ by a series of local operators
(for renormalizable theories, by a finite "renormalization" of the
lagrangian) that could be absorbed into $\delta\mathcal{L}
(\cdots;\{\sigma\})$:$$\delta\mathcal{L}(\cdots;\{\sigma\})
=\delta_{\text{var}}\mathcal{L}_{\text{QED}}+\tilde{\delta}
\mathcal{L}(\cdots;\{\sigma\}),\quad\delta_{\text{var}}
\mathcal{L}_{\text{QED}}\equiv\tilde{\mathcal{L}}_{\text{QED}}
-\mathcal{L}_{\text{QED}}.$$ Obviously, any sensible prescription of
$\breve{\mathcal{P}}_{\texttt{\scriptsize LE}}[\mathcal{L}
_{\text{eff}}]$ should necessarily include a reference scale that
are widely separated from the underlying ones. This natural
appearance of a reference scale just corresponds to and hence remove
the mystery around the dimensional transmutation phenomenon in the
conventional programs.
\subsection{Origin of renormalization group equations} We note that
renormalization group (RG) and RG equations (RGE) appear in terms of
intermediate renormalization reparametrization, for which a lucid
discussion can be found in Ref.\cite{muta}. If we could proceed with
all parameters being physical, there seems to be no room for it.
However, as discussed above, even starting with a physical
parametrization, one could still arrive at an arbitrary
prescription, provided the prescription is related to the physical
one via proper transformations. In this short subsection, we will
not repeat the derivation of RGE in a specific scheme defined by the
corresponding boundary conditions as is done in conventional
programs, instead, we briefly show that the RGE like equations could
be generically derived as a "decoupling" theorem in terms of
underlying structures even if the parameters are determined by
physical boundary conditions\cite{PLB6251,PLB6252,PLB6253}.

First let us look at the scaling law of QED in terms of underlying
theory. For any 1PI $n$-point vertex function $\Gamma^{(n)}$, the
scaling law would read,\bea\Gamma^{(n)}([\lambda p],[\lambda m,e];
\{\lambda^{d_{\sigma}}\sigma\})=\lambda^{d_{\Gamma^{(n)}}}
\Gamma^{(n)}([p],[m,e];\{\sigma\}),\eea with $d_{\cdots}$ denoting
the canonical scaling dimension (equals to the mass dimension) of
the corresponding parameter. The differential form of this scaling
law reads,\bea\left\{\sum_pp\cdot\partial_p+m\partial_m
+\sum_{\sigma}d_{\sigma}\sigma\partial_{\sigma}-d_{\Gamma^{(n)}}
\right\}\Gamma^{(n)}([p],[m,e];\{\sigma\})=0.\eea Now applying the
"decoupling" limit to this differential equation, we have,\bea
\label{diffscale}\left\{\sum_pp\cdot\partial_p+m\partial_m+\sum_{i}
d_{C_i}C_i\partial_{C_i}-d_{\Gamma^{(n)}}\right\}\Gamma^{(n)}
([p],[m,e]; [C_i])=0,\eea with\bea\sum_{i}d_{C_i}C_i\partial_{C_i}
\Gamma^{(n)}=\breve{\mathcal{P}}_{\texttt{\scriptsize LE}}\left(
\sum_{\sigma}d_{\sigma}\sigma\partial_{\sigma}\Gamma^{(n)}\right)
\eea denoting the contributions arising from the "decoupling" limit
operation.

Note that, in whatever prescription for the agent constants $[C_i]$,
the scaling of these constants would just elicit insertion of
appropriate local operators with respect to each loop where agent
constants show up: \bea\label{RGE-decoupling}\sum_{i}d_{C_i}C_i
\partial_{C_i}\Gamma^{(n)}=\sum_{a}\delta_{O_a}\check{I}_{O_a}
\Gamma^{(n)},\eea where $\check{I}_{O_a}$ denotes the insertion of a
local operator $O_a$ that corresponds to a vertex whose 1PI loop
corrections contain local ambiguities with $\delta_{O_a}$ being the
associated coefficients. At least at one-loop level in QED, only
logarithmic ambiguities will be really contributing to these
"decoupling" effects that are anomalous in terms of the canonical
contents of quantum field theories. Thus, $\delta_{O_a}$ is the
primitive "anomalous dimension" of operator $O_a$. A closer analysis
will tell that among the operators $[O_a]$, there must be kinetic
ones for which the above variations would induce an extra finite
"renormalization" of certain field operators, the rest must be
characterizing interactions for which the variations would give rise
to beta functions of their couplings\cite{PLB6251,PLB6252,PLB6253}.
So, Eq.(\ref{RGE-decoupling}) is our primitive form of
renormalization group equations for a general prescription, which
could be naturally interpreted as a "decoupling theorem" of the
underlying structures that regulate the ultra-violet regions of our
quantum field theories. Obviously, imposing different boundary
conditions would lead to different form of $\sum_{i}d_{C_i}C_i
\partial_{C_i}$ and $\sum_{a}\delta_{O_a}\check{I}_{O_a}$.

In fact, one could recast the scaling law of Eq.(\ref{diffscale})
into the following concise form\bea\{p\cdot\partial_p-d_{
\Gamma^{(n)}}\}\Gamma^{(n)}([p],[m,e];[C_i])&&={\hat{I}}_
{i\tilde{\Theta}}\Gamma^{(n)}([p],[m,e];[C_i])\nonumber\\&&=
\Gamma^{(n)}_{i\tilde{\Theta}}([0;p],[m,e];[C_i]), \eea with
${\hat{I}}_{i\tilde{\Theta}}\equiv-m\partial_m-\sum_{i}d_{C_i}
C_i\partial_{C_i}$ denoting the insertion of the full trace of the
stress tensor $\tilde{\Theta}^{\mu\nu}$ that contains trace
anomalies due to renormalization. This equation is just an
alternative form of the Callan-Symanzik equation, from which some
low-energy theorems of QCD follow as immediate
corollaries\cite{JPA40}.
\section{Discussions and summary}
Before closing our presentation, we wish to make the following
remarks about our simple strategy for renormalization: (1) As we
introduce no artificial deformation of all the canonical structures
or symmetries, hence not affecting spacetime dimension and Dirac
algebra, thus our strategy could be applied to any field theory,
such as supersymmetric and chiral field theories. (2) Since our
ignorance about the underlying theory is parametrized in a general
manner, the effects of regularization upon field theory and
corresponding physics could be examined in a general manner. In
particular, the physical origin of field theoretical properties like
anomalies could be unambiguously identified and disentangled with
the regularization effects in this strategy, surpassing the
traditional interpretations that rely on effects of regularization.
(3) Various operations such as external momenta routing and shift of
loop variables are allowed in our simple strategy, sparing many
subtleties associated these operations in the conventional
approaches. Hence, one could better focus on more physical issues in
calculations or identify true physical origins of various phenomena
within field theoretical contexts. (4) In theoretical perspective,
our approach is also useful in exploring whether or to what extent
the underlying structures are compatible with the canonical
symmetries or properties of quantum field theories, such as Lorentz
invariance, various gauge symmetries, unitarity and so on. From a
more practical viewpoint, our approach or similar strategies allows
us to examine how our ignorance about underlying structures could be
safely treated with the helps of these canonical symmetries.
Obviously, more works need to be done both in the construction and
applications of efficient programs using the simple strategy and in
the exploration of various issues that are intricate with respect to
the issue of regularization and renormalization.

In summary, we demonstrated in details with QED at one-loop level
how a simple strategy for renormalization could work without
introducing any specific form of regularization and manipulations of
ultra-violet infinities. Some technical operations like loop
variable shifting or momenta routing that are subtle issues in
conventional programs are shown to be of no problem. In this simple
strategy, ambiguities arise instead of infinities, which could be
further reduced by imposing appropriate Ward identities. For the
Lorentz and gauge invariant QED at one-loop level, it is shown in a
prescription-independent manner that there are only three
ambiguities to be fixed using appropriate boundary conditions. The
conventional renormalization programs were also analyzed from the
underlying theory perspective with the rationale of manipulation of
ultra-violet infinities being explicated. Finally, the
renormalization-group-like equations arise generically as
"decoupling theorems" of the underlying structures.
\section*{Acknowledgement}The project is supported in part by the
National Natural Science Foundation under Grant No.10205004 and the
Ministry of Education of China.
\appendix
\section{}
In this appendix, we list the convergent integrals that appear in
the RHS of Eqs.(\ref{diffsf},\ref{diffvt},\ref{diffvp}),\bea
-i\Sigma^{(1); \alpha\beta}(p)=&&(-ie)^2\int\frac{d^4l}{(2\pi)^4}
\frac{i\left[(1-\xi)\frac{l^\kappa l^\tau}{l^2}-g^{\kappa\tau}
\right]}{l^2 +i\epsilon}\gamma_\kappa\frac{i}{\slash\!\!l
+\slash\!\!\!p-m+i\epsilon}\nonumber\\&&
\times\left[\gamma^\alpha\frac{-1}{\slash\!\!l+\slash\!\!\!p
-m+i\epsilon}\gamma^\beta +(\alpha\leftrightarrow\beta)\right]
\frac{-1}{\slash\!\!l+\slash\!\!\!p-m+i\epsilon}\gamma_\tau;\\
\Lambda_\mu^{(1);\alpha}(p,\tilde{p})=&&(-ie)^2\int\frac{d^4l}
{(2\pi)^4}\frac{i\left[(1-\xi)\frac{l^\kappa l^\tau}{l^2}
-g^{\kappa\tau}\right]}{l^2+i\epsilon}\gamma_\kappa\frac{i}
{\slash\!\!l+\slash\!\!\!p-m+i\epsilon}\gamma^\alpha\nonumber\\
&&\times\frac{-1}{\slash\!\!l+\slash\!\!\!p-m+i\epsilon}
\gamma_\mu\frac{i}{\slash\!\!l+\slash\!\!\!\tilde{p}-m+i\epsilon}
\gamma_\tau;\\i\Pi^{(1);\alpha\beta\gamma}_{\mu\nu}(q)=&&(-ie)^2
\int\frac{(-)d^4l}{(2\pi)^4}\texttt{Tr}\left[\gamma_\mu\frac{i}
{\slash\!\!l-m+i\epsilon}\gamma_\nu\frac{i}{\slash\!\!l+\slash
\!\!\!q-m+i\epsilon}\gamma^\alpha\right.\nonumber\\&&\times\left.
\frac{-1}{\slash\!\!l+\slash \!\!\!q-m+i\epsilon}\gamma^\beta
\frac{-1}{\slash\!\!l+\slash\!\!\!q-m+i\epsilon}\gamma^\gamma
\frac{-1}{\slash\!\!l+\slash\!\!\!q-m+i\epsilon}\right]\nonumber\\
&&+\texttt{permutations of} (\alpha,\beta,\gamma).\eea Obviously,
these diagrams result in from zero momentum insertion of photon
lines into the original diagrams that are ill-defined, Cf. Fig. 2
for a diagrammatic representation for the case of self-energy
diagram.

As stated in Sec. II and III, other choices of routing only lead to
equivalent results. For the self-energy diagram, we could let the
external momentum $p$ of fermion flow through the photon line, so
that the RHS of Eq.(\ref{diffsf}) reads,\bea-i\Sigma^{(1);
\text{alt}}(p)=(-ie)^2\int\!\!\frac{d^4l}{(2\pi)^4}\frac{i\left
[(1-\xi)\frac{(l-p)^\kappa(l-p)^\tau}{(l-p)^2}-g^{\kappa\tau}
\right]}{(l-p)^2+i\epsilon}\gamma_\kappa\frac{i}{\slash\!\!l-m
+i\epsilon}\gamma_\tau.\nonumber\\\eea It is nothing else but the
original one with the loop momentum shifted by $p$. Then applying
our method to this integration, we arrived the following,
\bea-i\Sigma^{(1);\text{alt}}(p)=&&\frac{ie^2}{(4\pi)^2}\left\{
\int_0^1\!\!dz\left[[(z(2-3z)+z(3z-4)\xi)\slash\!\!\!p+
(4+2z(\xi-1))m]\right.\right.\nonumber\\&&\left.\left.\times
\ln(m^2-zp^2)-(1-\xi)z(1-z)\frac{p^2(m+z\slash\!\!\!p)}{m^2-zp^2}
\right]+\tilde{C}_\psi\slash\!\!\!p+\tilde{C}_m\right\},
\nonumber\\\eea which is equivalent to the form of the self-energy
diagram as given in Eq.(\ref{solutionsf}) after carrying out the
parametric integrations, and the constants may differ from the those
in Eq.(\ref{solutionsf}) by a finite amount. Indeed, computing the
difference between this routing and that given in
Eq.(\ref{solutionsf}), we have,\bea -i(\Sigma^{(1);\text{alt}}
-\Sigma^{(1)})=\frac{ie^2}{(4\pi)^2}\left\{\left(\tilde{C}_\psi
-C_\psi+\frac{(\xi-1)}{2}\right)\slash\!\!\!p +\tilde{C}_m-C_m
\right\}.\eea Thus the two routings produce identical result after
setting \bea\tilde{C}_\psi=C_\psi+\frac{(1-\xi)}{2},\quad
\tilde{C}_m=C_m, \eea which is natural from the underlying theory
perspective where we should have $-i(\Sigma^{(1);\text{alt}}
-\Sigma^{(1)})=0$. Of course, one could also verify this point using
a consistent regularization method like dimensional scheme. This is
an illustration of the theorem of routing in case (a) given in Sec.
II.

Alternatively, we could proceed as follows: First, we could find the
following from what we obtained above\bea-i\Sigma^{(1)}\|_{p=0}=&&
\frac{ie^2}{(4\pi)^2}\left[C_m+(3+\xi)m\ln m^2\right],\\-i
\Sigma^{(1);\text{alt}}\|_{p=0}=&&\frac{ie^2}{(4\pi)^2}\left[
\tilde{C}_m+(3+\xi)m\ln m^2\right].\eea Next note that,\bea-i
\Sigma^{(1)}\|_{p=0}=&&(-ie)^2\int\frac{d^4l}{(2\pi)^4}\frac{i
\left[(1-\xi)\frac{l^\kappa l^\tau}{l^2}-g^{\kappa\tau}\right]}
{l^2+i\epsilon}\gamma_\kappa\frac{i}{\slash\!\!l-m+i\epsilon}
\gamma_\tau\nonumber\\=&&-i\Sigma^{(1);\text{alt}}\|_{p=0},\eea
which implies $C_m=\tilde{C}_m$. In the same fashion, we could find
that\bea -i\partial_{p^\alpha}\Sigma^{(1)}\|_{p=0}=&&\frac{ie^2}
{(4\pi)^2}\left[C_\psi-\xi\ln m^2+\frac{1-\xi}{2}\right]
\gamma_\alpha,\\-i\partial_{p^\alpha}\Sigma^{(1);\text{alt}}\|_{p=0}
=&& \frac{ie^2} {(4\pi)^2}\left[\tilde{C}_\psi-\xi\ln m^2\right]
\gamma_\alpha.\eea Then from underlying theory perspective the two
quantities $-i\partial_{p^\alpha}\Sigma^{(1)}\|_{p=0}$ and
$-i\partial_{p^\alpha}\Sigma^{(1);\text{alt}}\|_{p=0}$ should be
identical and hence $\tilde{C}_\psi=C_\psi+\frac{(1-\xi)}{2}$. In
fact, the integral forms of these two quantities look different from
each other, but it could be verified in any consistent
regularization (e.g., dimensional regularization) that their
difference is zero, \bea-i\partial_{p^\alpha}(\Sigma^{(1)}
-\Sigma^{(1);\text{alt}})\|_{p=0}&&=\int\frac{d^4l}{(2\pi)^4}
\frac{e^2(3-\xi)\gamma^\beta} {l^2(l^2-m^2)}\left[g_{\beta\alpha}
-\frac{2l_\beta l_\alpha}{l^2}-\frac{2l_\beta l_\alpha}{l^2-m^2}
\right]\nonumber\\&&=0,\eea therefore they are identical in any
consistent regularization scheme (including the postulated
underlying theory), and the case (a) of the theorem of routing in
Sec. II is verified.

We also note that the routing for the vertex diagram given above are
so chosen that the calculations are less laborious, i.e., the two
external fermion lines carry independent momenta so that the
differentiation with respect to $p$ only operates on one internal
line. One could well try the following more conventional routing,
\bea\Lambda_\mu^{(1);\alpha}(p,p+q)=&&(-ie)^2\int\frac{d^4l}
{(2\pi)^4}\frac{i\left[(1-\xi)\frac{l^\kappa l^\tau}{l^2}
-g^{\kappa\tau}\right]} {l^2+i\epsilon}\gamma_\kappa\frac{i}
{\slash\!\!l+\slash\!\!\!p-m+i\epsilon}\nonumber\\&&\times\left[
\gamma^\alpha\frac{-1}{\slash\!\!l+\slash\!\!\!p-m+i\epsilon}
\gamma_\mu+\gamma_\mu\frac{-1}{\slash\!\!l+\slash\!\!\!p+\slash
\!\!\!q-m+i\epsilon}\gamma^\alpha\right]\nonumber\\&&\times
\frac{i}{\slash\!\!l+\slash\!\!\!p+\slash\!\!\!q-m+i\epsilon}
\gamma_\tau.\eea The final results for $\Lambda_\mu$ should be the
same after we set $q=\tilde{p}-p$ in the solutions. After quite some
algebras, we found that the vertex computed using this routing
reads, \bea\Lambda^{(1);\alpha}_\mu(p,p+q)=&&\partial_{p_\alpha}
\left\{-\frac{e^2}{(4\pi)^2}\int_0^1\!\!\!dy\int_0^{1-y}\!\!\!dz
\left[2\gamma_\mu\ln\tilde{\Delta}+\frac{\tilde{\mathcal{G}}_\mu}
{\tilde{\Delta}}+(1-\xi)(1-y-z)\right.\right.\nonumber\\&&\left.
\left.\times\left(-6\gamma_\mu\ln \tilde{\Delta}+\frac{\tilde
{\mathcal{F}}_{1;\mu}}{\tilde{\Delta}^2}+\frac{\tilde{\mathcal{F}}
_{2;\mu}} {\tilde{\Delta}}\right)\right]\right\},\nonumber\\
\tilde{P}\equiv &&(y+z)p+yq,\ \tilde{\Delta}\equiv(y+z)m^2
+\tilde{P}^2-zp^2 -y(p+q)^2,\nonumber\\
\tilde{\mathcal{G}}_\mu\equiv &&\gamma^\sigma(\slash\!\!\!p+m
-\slash\!\!\!\tilde{P})\gamma_\mu(\slash\!\!\!p+\slash\!\!\!q
+m-\slash\!\!\!\tilde{P})\gamma_\sigma,\nonumber\\
\tilde{\mathcal{F}}_{1;\mu}\equiv&&\slash\!\!\!\tilde{P}(\slash
\!\!\!p+m-\slash\!\!\!\tilde{P})\gamma_\mu(\slash\!\!\!p+\slash
\!\!\!q+m-\slash\!\!\!\tilde{P})\slash\!\!\!\tilde{P},\nonumber\\
\tilde{\mathcal{F}}_{2;\mu}\equiv&&(\slash\!\!\!p+\slash\!\!\!q)
\gamma_\mu\slash\!\!\!p+3\gamma_\mu(\slash\!\!\!p+\slash\!\!\!q)
\slash\!\!\!\tilde{P}+3\slash\!\!\!\tilde{P}\slash\!\!\!p\gamma_\mu
+(m^2-6\tilde{P}^2)\gamma_\mu\nonumber\\&&+2m(3\tilde{P}_\mu-2p_\mu
-q_\mu).\eea It is immediate to see that this result exactly agrees
with that given above after setting $\tilde{p}=p+q$, as asserted by
the theorem of routing in case (b) presented in Sec. II.

\section{}
Here, we consider the Ward identity for the components proportional
to $1-\xi$ that are listed as below:\bea -i\frac{\Delta\Sigma^{(1)}
(p;\epsilon)}{1-\xi}&&=-\frac{ie^2}{(4\pi)^2}\left\{\left[
\epsilon^{-1}-\gamma_E+1-\ln\frac{m^2}{4\pi\mu^2}+\delta
+(\delta^2-1)\ln\frac{\delta-1}{\delta}\right ]\slash\!\!\!p\right.
\nonumber\\&&\quad\left.-\left[\epsilon^{-1}-\gamma_E+2-\ln\frac{
m^2}{4\pi\mu^2}+(\delta-1)\ln\frac{\delta-1}{\delta}\right]m\right
\}\nonumber\\&&=-\frac{ie^2}{(4\pi)^2}\left\{\left[\frac{\Delta
C_\psi(\epsilon)}{\xi-1}-\ln m^2+\delta+(\delta^2-1)\ln\frac{
\delta-1}{\delta}\right]\slash \!\!\!p\right.\nonumber\\&&\quad
\left.-\left[\frac{\Delta C_m(\epsilon)}{(1-\xi)m}+1-\ln m^2
+(\delta-1)\ln\frac{\delta-1}{\delta}\right]m\right\},\\
\frac{\Delta\Lambda^{(1)}_\mu(p,p;\epsilon)}{1-\xi}&&=-\frac{e^2}
{(4\pi)^2}\left\{\left[\epsilon^{-1}-\gamma_E+1-\ln\frac{m^2}
{4\pi\mu^2}+\delta+(\delta^2-1)\ln\frac{\delta-1}{\delta}\right]
\gamma_\mu\right.\nonumber\\&&\quad\left.-2\left[\left(\delta
\ln\frac{\delta-1}{\delta}+1\right)(2\delta\slash\!\!\!p-m)
+\slash\!\!\!p\right]\frac{p_\mu} {p^2}\right\}\nonumber\\&&
=-\frac{e^2}{(4\pi)^2}\left\{\left[\frac{\Delta C_\perp
(\epsilon)}{1-\xi}+\frac{11}{6}-\ln m^2+\delta+(\delta^2-1)
\right.\right.\nonumber\\&&\left.\left.\ \ \times\ln\frac{\delta-1}
{\delta}\right]\gamma_\mu-2\left[\left(\delta\ln\frac{\delta-1}
{\delta}+1\right)(2\delta\slash\!\!\!p-m)+\slash\!\!\!p\right]
\frac{p_\mu}{p^2} \right\},\eea with $\delta\equiv m^2/p^2$.
Differentiating $\Delta\Sigma^{(1)}(p;\epsilon)$ with respect to
$p^\mu$, we have, \bea-i\partial_{p^\mu}\left\{\frac{\Delta
\Sigma^{(1)}(p;\epsilon)}{1-\xi}\right\}&&=-\frac{ie^2}{(4\pi)^2}
\left\{\left[\epsilon^{-1}-\gamma_E +1-\ln\frac{m^2}{4\pi\mu^2}
+\delta +(\delta^2-1)\right.\right.\nonumber\\&&\left.\left.\
\times\ln \frac{\delta-1} {\delta}\right]\gamma_\mu-2\left[\left(
\delta\ln\frac{\delta-1}{\delta}+1\right)(2\delta\slash\!\!\!p-m)
+\slash\!\!\!p\right]\frac{p_\mu}{p^2}\right\}\nonumber\\
&&=i\frac{\Delta\Lambda^{(1)}_\mu(p,p;\epsilon)}{1-\xi}.\eea
Obviously, the Ward identity for the components proportional to
$(1-\xi)$ is guaranteed by the following relation between $\Delta
C_\psi(\epsilon)$ and $\Delta C_\perp(\epsilon)$:\bea\Delta C_\psi
(\epsilon)+\Delta C_\perp(\epsilon) =\frac{11(\xi-1)}{6}.\eea Now,
collecting all the components for $C_\psi(\epsilon)$ and
$C_\perp(\epsilon)$:
$$C_\psi(\epsilon)=C_\psi(\epsilon)\|_{\xi=1}+\Delta C_\psi
(\epsilon),\quad C_\perp(\epsilon)=C_\perp(\epsilon)
\|_{\xi=1}+\Delta C_\perp(\epsilon),$$ we have:\bea C_\psi
(\epsilon)+C_\perp(\epsilon)=\frac{5}{2}+\frac{11(\xi-1)}
{6}=\frac{4+11\xi}{6}.\eea

In addition, we note that in Landau gauge,\bea-i\Sigma^{(1)}
(p;\epsilon)\|_{\xi=0}&&=-\frac{3ie^2}{(4\pi)^2}\left[\epsilon^{-1}
-\gamma_E+\frac{4}{3}-\ln\frac{m^2}{4\pi\mu^2}\right.\nonumber\\&&
\left.\quad+(\delta-1)\ln\frac{\delta-1}{\delta}\right]m,\\
\Lambda^{(1)}_\mu(p,p;\epsilon)\|_{\xi=0}&&=\frac{6e^2}{(4\pi)^2}
\left[\delta\ln\frac{\delta-1}{\delta}+1\right]\frac{mp_\mu}{p^2},
\eea as it is well known that the vertex is ultra-violet definite in
this gauge.

\section{}
The differentiation of other diagrams with respect to $p_{1;\alpha}$
yields,\bea\partial_{p_{k;\alpha}}i\Gamma_{k;\mu
\nu\rho\sigma}=&&\left(\frac{8ie^4}{(4\pi)^2}T_{k;\mu\nu\rho\sigma}
\int_0^1\!\!\!dx \int_0^x\!\!\!dy\int_0^y\!\!\!dz\partial_{p_{1;
\alpha}}\ln\tilde{\Delta}_k\right)\nonumber\\&&
+\partial_{p_{1;\alpha}}\check{\Gamma}_{k;\mu\nu\rho\sigma},\quad
k=2,3,4,5,6,\eea with\bea&&T_{2;\mu\nu\rho\sigma}=g_{\mu\rho}
g_{\nu\sigma}-2g_{\mu\nu}g_{\rho\sigma}+g_{\mu\sigma}g_{\nu\rho},
\nonumber\\&&T_{3;\mu\nu\rho\sigma}=g_{\mu\nu}g_{\rho\sigma}
-2g_{\mu\sigma}g_{\nu\rho}+g_{\mu\rho}g_{\nu\sigma},\nonumber\\
&&T_{4;\mu\nu\rho\sigma}=g_{\mu\sigma}g_{\nu\rho}-2g_{\mu\rho}
g_{\nu\sigma}+g_{\mu\nu}g_{\rho\sigma},\nonumber\\&&
T_{5;\mu\nu\rho\sigma} =g_{\mu\sigma}g_{\nu\rho}-2g_{\mu\nu}
g_{\rho\sigma}+g_{\mu\rho}g_{\nu\sigma},\nonumber\\
&&T_{6;\mu\nu\rho\sigma}=g_{\mu\rho}g_{\nu\sigma}-2g_{\mu\sigma}
g_{\nu\rho}+g_{\mu\nu}g_{\rho\sigma},\\&&\tilde{\Delta}_2\equiv
m^2-(x-x^2)p_1^2-(z-z^2)p_2^2-(y-y^2)p_3^2\nonumber\\&&\quad\quad\ \
-2z(1-x)p_1p_2-2y(1-x)p_1p_3-2z(1-y)p_2p_3,\nonumber\\
&&\tilde{\Delta}_3\equiv m^2-(x-z-(x-z)^2)p_1^2-(y-z-(y-z)^2)
p_2^2-(z-z^2)p_3^2\nonumber\\&&\quad\quad\ \ -2(y-z)(1-x+z)
p_1p_2-2z(x-z)p_1p_3-2z(y-z)p_2p_3,\nonumber\\&&\tilde{\Delta}_4
\equiv m^2-(x-y-(x-y)^2)p_1^2-(y-y^2)p_2^2-(y-z-(y-z)^2)p_3^2\nonumber\\
&&\quad\quad\ \ -2y(x-y)p_1p_2-2(x-y)(y-z)p_1p_3-2(y-z)(1-y)p_2p_3,
\nonumber\\&&\tilde{\Delta}_5\equiv m^2-(x-y-(x-y)^2)p_1^2
-(y-z-(y-z)^2)p_2^2-(y-y^2)p_3^2\nonumber\\&&\quad\quad\ \
-2(x-y)(y-z)p_1p_2-2y(x-y)p_1p_3-2(y-z)(1-y)p_2p_3,\nonumber\\&&
\tilde{\Delta}_6\equiv m^2-(x-z-(x-z)^2)p_1^2-(z-z^2)p_2^2
-(y-z-(y-z)^2)p_3^2\nonumber\\&&\quad\quad\ \ -2z(x-z)p_1p_2
-2(y-z)(1-x+z)p_1p_3-2z(y-z)p_2p_3.\eea

Thus the solutions:\bea i\Gamma_{k;\mu\nu\rho\sigma}=&&\frac{8ie^4}
{(4\pi)^2}T_{k;\mu\nu\rho\sigma}\int_0^1\!\!\!dx\int_0^x\!\!\!dy
\int_0^y\!\!\!dz\ln\frac{\tilde{\Delta}_k}{C_{k;4\gamma}}
\nonumber\\&&+\check{\Gamma}_{k;\mu\nu\rho\sigma},\quad
k=2,3,4,5,6,\eea with the same convention of notations. Here we have
deliberately denote the integration constants in the way that seems
diagram-dependent, i.e., $C_{;4\gamma}$ may differ among the six
diagrams. The gauge invariance will require that they must be equal
to each other. One could also arrive at the same conclusion by
noting that the loop integral for each diagram is identical after
setting all the external momenta to zero, for example, \bea
i\Gamma_{1;\mu\nu\rho\sigma}(0,0,0,0)=-e^4\texttt{Tr}[
\gamma_\alpha\gamma_\mu\gamma_\beta\gamma_\nu\gamma_\delta
\gamma_\rho\gamma_\eta\gamma_\sigma]\int\frac{d^4l}{(2\pi)^4}
\frac{l^\alpha l^\beta l^\delta l^\eta}{(l^2-m^2)^4}.\eea

\section{}
Now we demonstrate the integration by parts on the logarithmic term
$\ln\tilde{\Delta}_1$:\bea&&\int_0^1\!\!\!dx\int_0^x\!\!\!dy
\int_0^y\!\!\!dz\ln\frac{\tilde{\Delta}_1}{C_{1;4\gamma}}
\nonumber\\=&&\int_0^1\!\!\!dx\int_0^x\!\!\!dy\left[\left(z\ln
\frac{\tilde{\Delta}_1}{C_{1;4\gamma}}\right)\|_{z=0}^{z=y}
-\int_0^y\!\!\!dz\frac{z\partial_z\tilde{\Delta}_1}{\tilde{
\Delta}_1}\right]\nonumber\\=&&\int_0^1\!\!\!dx\int_0^x\!\!\!
dyy\ln\frac{\tilde{\tilde{\Delta}}_1}{C_{1;4\gamma}}+\tilde{D}_{1z}
\nonumber\\=&&\int_0^1\!\!\!dx\left[\left(\frac{y^2}{2}\ln\frac{
\tilde{\tilde{\Delta}}_1}{C_{1;4\gamma}}\right)\|_{y=0}^{y=x}
-\int_0^x\!\!\!dy \frac{y^2}{2}\frac{\partial_y\tilde{\tilde{
\Delta}}_1}{\tilde {\tilde{\Delta}}_1}\right]+\tilde{D}_{1z}
\nonumber\\=&&\int_0^1\!\!\!dx\frac{x^2}{2}\ln\frac{m^2-(x-x^2)
p_4^2}{C_{1;4\gamma}}+\tilde{D}_{1y}+\tilde{D}_{1z}\nonumber\\
=&&\frac{1}{6}\ln\frac{m^2}{C_{1;4\gamma}}+\tilde{D}_{1x}
+\tilde{D}_{1y}+\tilde{D}_{1z}=\frac{1}{6}\ln\frac{m^2}
{C_{1;4\gamma}}+\tilde{D}_{1},\eea with\bea&&\tilde{\tilde{
\Delta}}_1\equiv \tilde{\Delta}_1\|_{z=y},\quad\tilde{D}_{1z}
\equiv-\int_0^1\!\!\!dx\int_0^x\!\!\!dy\int_0^y \!\!\!dz
\frac{z\partial_z\tilde{\Delta}_1}{\tilde{\Delta}_1},\nonumber\\&&
\tilde{D}_{1y}\equiv-\int_0^1\!\!\!dx\int_0^x\!\!\!dy\frac{y^2}{2}
\frac{\partial_y\tilde{\tilde{\Delta}}_1}{\tilde{\tilde{
\Delta}}_1},\quad\tilde{D}_{1x}\equiv\frac{1}{6}\int_0^1\!\!\!
dx\frac{x^3(1-2x)}{x^2-x+m^2/p^2_4},\nonumber\\&&\tilde{D}_1
\equiv\tilde{D}_{1x}+\tilde{D}_{1y}+\tilde{D}_{1z}.\eea Here
$\tilde{D}_1$ is definite and hence absorbed into $\check{\Gamma}_1$
in Eq.(\ref{4gamma-repres}), that is, $\tilde{\Gamma}_1\equiv
\check{\Gamma}_1+\tilde{D}_1$. Then we arrive at the form of
$\Gamma_{1;\mu\nu\rho\alpha}$ given in Eq.(\ref{4gamma-repres}). In
the same fashion, we have,\bea\int_0^1\!\!\!dx\int_0^x\!\!\!dy
\int_0^y\!\!\!dz\ln\tilde{\Delta}_k=\frac{1}{6}\ln\frac{m^2}
{C_{k;4\gamma}}+\tilde{D}_{k},\quad k=2,3,4,5,6,\eea with
$[\tilde{D}_{k}, k=2,3,4,5,6]$ being definite functions of external
momenta.

\vspace{2cm}
\begin{figure}[h]
\centerline{\psfig{file=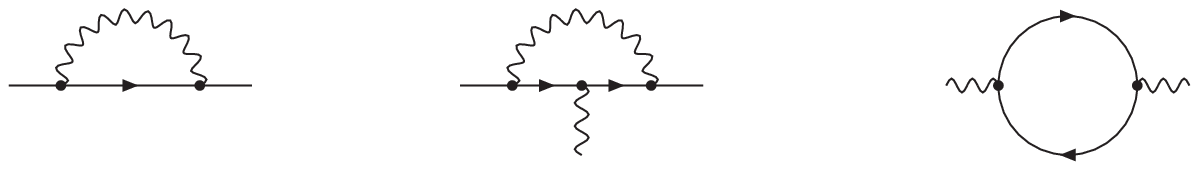,width=15cm}}\vspace{-1cm}
\caption{The elementary one-loop diagrams in QED}
\end{figure}
\begin{figure}[h]\hspace{0.3cm}
\centerline{\psfig{file=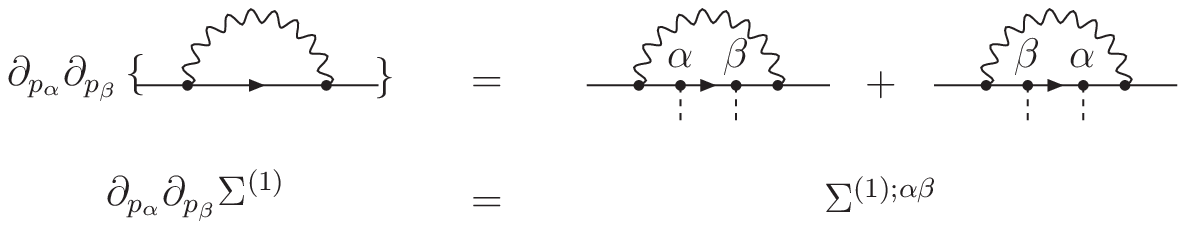,width=15cm}}\vspace{-0.5cm}
\caption{The graphical representation of Eq.(14).}
\end{figure}
\end{document}